\documentclass{aa}  

\usepackage{graphicx}
\usepackage{txfonts}
\usepackage{url}
\usepackage[breaklinks=true]{hyperref}
\usepackage{natbib} 
\bibpunct{(}{)}{;}{a}{}{,} 

\usepackage{verbatim}

\usepackage{epstopdf}

\begin{document}
   \title{Evolution of the habitable zone 
     of low-mass stars}

   \subtitle{Detailed stellar models and analytical relationships for
     different masses and chemical compositions
\thanks{http://astro.df.unipi.it/stellar-models/HZ/}
}

   \author{G. Valle \inst{1,2}, M. Dell'Omodarme \inst{1}, P.G. Prada Moroni
     \inst{1,2}, S. Degl'Innocenti \inst{1,2}
          }

   \authorrunning{Valle, G. et al.}

   \institute{Dipartimento di Fisica ``Enrico Fermi'',
Universit\`a di Pisa, largo Pontecorvo 3, Pisa I-56127 Italy
\and
  INFN,
 Sezione di Pisa, Largo B. Pontecorvo 3, I-56127, Italy}

   \offprints{G. Valle, valle@df.unipi.it}

   \date{Received 28/12/2013; accepted 23/05/2014.}

  \abstract
   {  
The habitability of an exoplanet is assessed by determining the times at which its
orbit lies in 
the circumstellar habitable zone (HZ). This zone evolves with time following
the stellar luminosity variation, which means that the time spent in the HZ depends on the
evolution of the host star.   
}
   {  
We study the temporal evolution of the HZ of low-mass stars --
only due to stellar evolution --  
and evaluate the related uncertainties.
These uncertainties are then compared with those due to
the adoption of different climate models.
}
{  
We computed stellar evolutionary tracks from the pre-main sequence
phase 
to the helium flash at the red-giant branch tip for stars with masses in the
range [0.70 - 1.10] $M_{\sun}$, metallicity $Z$ in the range
[0.005 - 0.04], and various initial helium contents. By adopting a reference
scenario for the HZ computations, we
evaluated several characteristics of the HZ, such as the distance from the host
star at which the habitability is longest, the duration of this
habitability, the width of the zone for which the habitability lasts one half
of the maximum, and the boundaries of the continuously habitable zone (CHZ)
for which the habitability lasts 
at least 4 Gyr. We developed analytical models, accurate to the percent level
or lower,
which allowed to obtain these characteristics in dependence on the mass and the chemical
composition of the host star. 
}
  {
The metallicity of the host star plays a
relevant role in determining the HZ. The importance of the initial helium
content  is evaluated here for the first time; it accounts for
a variation of the CHZ boundaries as large as 30\% and 10\% in the inner and
outer border. 
The computed analytical models allow the first systematic study of the
variability  
of the CHZ boundaries that is caused by the uncertainty in the estimated
values of mass 
and metallicity of the host star. An uncertainty range of about 30\% in the
inner  
boundary and 15\% in the outer one were found. 
We also verified that these uncertainties are larger than that due to relying
on recently revised climatic 
  models, which leads to a CHZ boundaries shift within $\pm5\%$ with respect
  to those of our reference scenario. 
 We made an on-line tool available that provides both HZ characteristics and
interpolated stellar tracks.
 }
{}


   \maketitle

\section{Introduction}\label{sec:intro}

In the light of our limited knowledge, the existence of life is strongly
associated 
with the presence of liquid water. 
The habitable zone (HZ) is then defined as the range of distances
from a star within which a planet may contain liquid water at its
surface \citep[see
  e.g.][]{Hart1978,
  Kasting1993,Underwood2003,Selsis2007,Pierrehumbert2011,Kopparapu2013}.   
Several potentially habitable planets are already reported in literature
\citep[see among others][]{Udry2007, Pepe2011, Borucki2011, Borucki2012,
  Anglada2012, Barclay2013, Tuomi2013, Batista2014}, and their number is
expected to increase with time.   

The position of the HZ around the host star is not immutable with time, because
it, and its width, are influenced by different climatological, biogeochemical,
geodynamical, and astrophysical processes \citep[see e.g.][]{Kasting1993,
  Forget1998, Selsis2007, vonBloh2009, Kopparapu2013, Leconte2013}.  
Although it is expected that -- following the advance in the
understanding of the interaction of all these processes --  complex HZ models will
eventually account self-consistently for all 
these aspects, at present no such model is available. 
 
In this paper we focus on the evolution of the position of
  the HZ
boundaries due only to the luminosity change of the host star.
As explained in detail in Sect.~\ref{sec:methods}, we evaluate some relevant
HZ characteristics and quantify 
 the uncertainty that affects these estimates only originating
from the stellar observational uncertainties. To set these uncertainties in
context, we also compare these values
to those caused by different approximations in the HZ computations.

While evaluating the stellar uncertainty contribution to the 
overall uncertainty budget is only a single aspect of a very complex problem, at
present it  
represents one of the firmest parts from the point of view of both 
observation and theory. Concerning the former, the stellar observables, 
such as luminosity, effective temperature, and metallicity can be 
determined much more accurately than the planetary ones. 
Concerning the latter, stellar evolution theory is one of the 
most mature and tested in the astrophysical literature. 
Moreover, the stellar impact on the HZ evolution will remain an essential 
ingredient in the future as well, when a more complete approach will be developed.

We
restrict our analysis to low-mass stars -- [0.70 - 1.10]~$M_{\sun}$ -- for a
large set of metallicities $Z$ -- [0.005 - 0.04] -- and initial helium
abundances $Y$.  For more 
massive stars a similar study is reported in
\citet{Lopez2005} and \citet{Danchi2013}.

To evaluate the boundaries of the HZ, we relied on 
effective stellar fluxes obtained by climate models (see
  Sect.~\ref{sec:methods}). In our HZ reference scenario computations we also
  assumed a constant planetary albedo and adopted the equilibrium temperature
-- that is, the temperature of a planet in thermal equilibrium between
the radiation absorbed from the star and that radiated into space --
  as the 
  parameter defininig the position of the HZ. 
 Within this scenario, the temporal  
  evolution of the temperature of a planet at a given distance from the 
  host star only depends on the evolution of the stellar luminosity. 
  The boundaries of the HZ will then evolve
following the changes of the luminosity with time.  A planet is considered
habitable if its equilibrium temperature is within the range of the allowed
temperature obtained by climate models.

At variance with the other works published in the literature, we also provide
here analytical relations of the dependence on the stellar mass and 
chemical composition of several important HZ features, namely the distance
from the host star at which the longest duration of habitability occurs, the
duration 
of this longest habitability, the width of the zone for which the habitability
lasts at least one half of the maximum, and the position of the inner and outer
boundaries for which the habitability lasts at least 4~Gyr.
These analytical models have three purposes. First, they allow a
systematic 
study of the variability of the HZ boundaries position due to the uncertainty
in the estimates of the host star mass and metallicity, which is still lacking
in the 
literature. 
Second, they allow distinguishing the contribution of the various
stellar characteristics to the temporal variation of the HZ.  Third, after
a HZ boundary scenario is chosen, these models show the high degree of
accuracy that can be achieved by analytical relations based on linear models.

We provide an on-line tool that allows a) downloading a stellar track for the
chosen  
mass and chemical composition and b) obtaining the HZ characteristics.

The structure of the paper is the following: in Sect.~\ref{sec:methods} we
discuss the method and the assumptions employed to computate the edges
of the HZ and provide details on the adopted stellar evolutionary code and on
the grid of the computed models.  The main results are reported in
Sect.~\ref{sec:results}, while the analytical models are
discussed in Sect.~\ref{sec:modelli}. In Sect.~\ref{sec:flux} we
 present a comparison of the results given in Sect.~\ref{sec:results} with
those obtained without the constant albedo assumption and using the models
by \citet{Kopparapu2013} (hereafter K13). 
Concluding remarks are presented in
Sect.~\ref{sec:conclusions}.

\section{Methods}\label{sec:methods}

We assumed an Earth-mass planet with an H$_2$O/CO$_2$/N$_2$
atmosphere.
Other atmospheric compositions, such as H$_2$ rich ones, would
change our results, moving the outer HZ
boundary farther from the host star \citep[see e.g.][]{Stevenson1999,
  Pierrehumbert2011}. We did not not consider this class
of objects here. 

The HZ boundaries can be computed by assuming a
grey-body approximation in which the radiative properties of the planet are
only described by the albedo $A$. The temporal variation of the HZ boundaries is
then described by the relation
\begin{equation}
d(t) \; ({\rm AU}) = \left( \frac{L(t)/L_{\sun}}{S_{\rm eff}} \right)^{1/2}\; ,
\label{eq:d-seff}
\end{equation}
where $L(t)/L_{\sun}$ is the stellar luminosity in solar units, and   
$S_{\rm eff}$ is the effective stellar flux 
defined as the value required to maintain a given surface
temperature and calculated by the climatic model as the ratio of the 
outgoing flux and the incident stellar flux. The contribution of the
planetary albedo, which is evaluated as a function of the effective
temperature of the 
host star, is internally accounted for in the models. 
These $S_{\rm eff}$ critical values are
determined for both a dense 
H$_2$O atmosphere at the inner edge and a dense CO$_2$ atmosphere with 1
  bar 
of N$_2$ and one earth gravity at the outer edge.

A common approach to the HZ distance computations is to assume a constant
planetary albedo \citep[see e.g.][]{Borucki2011,
  Batalha2013, Cantrell2013, Danchi2013}. 
Under this hypothesis, equating the stellar fluxes absorbed
and radiated by the planet \citep[see e.g.][for details of the
  computation]{Kaltenegger2011},  Eq.~(\ref{eq:d-seff}) can be rewritten as 
\begin{equation}
d(t) \; ({\rm AU}) = \frac{1}{a} \left( \frac{(1 - A) L(t)}{16 \pi \sigma}  \right)^{1/2}
\frac{1}{T_{\rm p}^2}\; ,
\label{eq:d-HZ}
\end{equation}
where $A$ is the Bond albedo, $\sigma$ is the Stefan-Boltzmann
constant, $a$ the conversion factor between centimeters and astronomic units,
and $T_{\rm p}$ is the planet equilibrium temperature. 
In this scenario, the inner ($d_{\rm i}$) and outer ($d_{\rm o}$) radii of the
HZ can be easily 
computed from Eq.~(\ref{eq:d-HZ}) after the albedo and the corresponding
temperatures $T_{\rm 
  i}$ and $T_{\rm o}$ are fixed.
The explicit link between effective stellar flux, albedo and equilibrium
  temperature is given by
\begin{equation}
T_{\rm p} = \left( \frac{S_{\rm eff} \; (1-A) \; L_{\sun}}{16 \pi \sigma \;
  a^2} \right)^{1/4} \; .
\end{equation}
In the computation we adopted Eq.~(\ref{eq:d-HZ}) as reference scenario, using
the Earth's average Bond albedo \citep[][i.e. 0.3]{dePater2001} for $A$. 


Significant progress is made in climate
modelling, due to the development of 3D models \citep[see e.g.][]{Leconte2013,
Leconte2013b}, which have shown systematic biases between mean
surface temperatures predicted by 1D and 3D
simulations. As a consequence, the associated estimates of the position of the
HZ 
boundaries are expected to change in the forthcoming years. 
Therefore, giving actual estimates of the HZ and HZ characteristics is not the
main goal of this paper. Instead, we are interested in analysing the
dependence of several chosen features of the HZ on stellar evolution and on the
quantification of the related uncertainties. This differential effect
is expected to be robust to a change of the boundary positions by 
improved climate models.   

We assumed as equilibrium temperature of the inner boundary of the
habitable zone $T_{\rm i}$~=~269~K \citep[see e.g.][]{Forget1998,Lopez2005,
  Selsis2007,Danchi2013}.  As discussed in detail in \citet{Selsis2007}, this
limit is determined considering that if the surface temperature remains below
the critical temperature of water (647 K), the thermal emission of an habitable
planet cannot exceed the runaway greenhouse threshold, equivalent to the
irradiance of a black-body at 270 K.  A planet with an atmosphere with an
equilibrium temperature above 270 K either has a surface temperature below 647
K, but without liquid water on its surface, or a considerable amount of water,
but a surface temperature above 1400 K. In both cases the planet would be
uninhabitable for life as we know it.

The definition of the outer boundary position is more complicated, since
  there is no well-accepted estimate 
for the lowest insulation compatible with stable liquid water under a dense
CO$_2$ atmosphere. To tackle this problem,   we adopted
a set of effective stellar fluxes from 0.4 to 0.2 \citep{Forget1997, Forget1998,
  Mischna2000}. These values  correspond to equilibrium temperatures
of $T_{\rm o}$~=~203~K  and $T_{\rm o}$~=~169~K, assuming an albedo of 0.3.
 The former is set by the lowest 
temperature at which the liquid-solid phase change of water can occur, which
depends on the existence of a greenhouse effect from CO$_2$ and H$_2$O
\citep[see e.g.][]{Kasting1993, Forget1998}. The latter is based on planetary
atmosphere models that include CO$_2$ ice clouds. Atmospheres with optically
thick CO$_2$ ice clouds with large particle radii can more easily maintain the
surface of 
a planet above the freezing point of water \citep{Forget1997, Mischna2000,
  Wordsworth2011}.

The adopted values of the inner and the outer borders equilibrium
  temperatures 
 and the albedo are the same of \citet{Danchi2013}, since we are 
  interested in a comparison of the overlapping results. This comparison
will give information about the uncertainty due only to the adoption of a
different set of 
stellar tracks in the HZ computations.

We defined as continuously habitable zone
(CHZ) a HZ lasting longer than 4~Gyr.
Other choices are possible and values from 1 to 5~Gyr are
adopted in  
literature \citep[see e.g.][]{Schopf1993, Turnbull2003, Buccino2006}.

Although the approach described above is
currently widely adopted, it has known limits. Its main shortcoming is the
assumption of an albedo 
$A$ independent of the spectrum of the incoming
radiation.  As a matter of 
fact, the Bond albedo is not a planetary characteristics alone, but it
also depends on the spectral energy distribution of the host star \citep[see
  e.g. the extensive discussion on][]{Selsis2007, Kaltenegger2011}.
Computations of HZ that do not assume a constant albedo are available in
literature 
\citep[e.g.][]{Selsis2007, Kopparapu2013}; but they strongly depend
 on the 
climatic models used for the planetary atmosphere, on its compositions, on the
scheme of cloud coverage, and on the considered geoclimatic processes, which
are related in a still unknown manner to the mass of the planet \citep[see
  e.g.][for details on these problems]{Selsis2007}.  
Since these factors controlling a planet surface temperature
are not yet constrained either by observations or by self-consistent models  
\citep[see][for a review]{Seager2013}, we preferred to adopt in this work the
simple  
model of Eq.~(\ref{eq:d-HZ}) as our reference scenario. 

However, it is interesting to evaluate
the effect on the results of excluding the constant albedo assumption.
In Sect.~\ref{sec:flux}
we compare the results of our reference scenario with those obtained using the
revised climate model presented in 
K13, without assuming a constant albedo.
A similar comparison is presented in \citet{Danchi2013} for more massive
stars. The correction is based on the parametrisation given in
 \citet{Selsis2007}. For the HZ characteristics considered here, 
the differences due to excluding the constant albedo assumption  
are
of the order of a few percent.
 
Figure~16 of \citet{Danchi2013} clearly shows that the duration of
 habitability of 
main-sequence solar and super-solar metallicity stars are essentially
unaffected by the 
refinements in the 
boundary positions. According to that figure, the changes only influence
the case at metallicity $Z$~=~0.001, which is below the range considered
here. Table~3 of the quoted paper 
shows that this statement also holds for the CHZ boundaries.

\subsection{Stellar evolution models}

The stellar models of this paper have been computed by means of
 the FRANEC \citep{scilla2008} evolutionary code. We used the most 
 updated version of the code, with the same input physics and parameters 
as were adopted to build the Pisa Stellar Evolution Data Base for low-mass 
 stars\footnote{\url{http://astro.df.unipi.it/stellar-models/}}
 \citep{database2012, stellar}.  
 We followed the evolution from the pre-main sequence phase to the helium
 flash at the tip of the red giant phase. Stellar tracks were computed in the
 mass range [0.70 - 1.10]~$M_{\sun}$, with a step of 0.05~$M_{\sun}$, for
 metallicity $Z$ in the range [0.005 - 0.04], with a step of 0.005. We adopted
 the solar heavy elements mixture by \citet{AGSS09}. As usually done in
 literature, the value of the initial helium abundance used as input parameter
 was obtained from the following linear relation:
\begin{equation}
Y = Y_p+\frac{\Delta Y}{\Delta Z} Z
\label{eq:YZ}
\end{equation}
with cosmological $^4$He abundance value $Y_p = 0.2485$, from WMAP
\citep{cyburt04,steigman06,peimbert07a,peimbert07b}, and assuming $\Delta
Y/\Delta Z = 2$ \citep{pagel98,jimenez03,gennaro10}. To account for the current
uncertainty on $\Delta Y/\Delta Z$, models with $\Delta Y/\Delta Z = 1$ and 3
were computed as well. We adopted our solar calibrated value of the mixing length
parameter, i.e. $\alpha_{\rm ml} = 1.74$. Further details
on the inputs adopted in the computations are available in \citet{cefeidi,
  incertezze1, incertezze2}. 
  
We computed 216 stellar evolutionary tracks specifically 
  for the present work.
  
Relying on this fine grid of detailed stellar models, we make available 
an on-line tool \footnote{\url{http://astro.df.unipi.it/stellar-models/HZ/}} 
that provides interpolated stellar tracks from the pre-main sequence 
to the red-giant branch tip for the required mass and chemical composition. 
The file provides as a function of time (in Gyr) the logarithm of stellar luminosity (in units
  of solar luminosity); the logarithm of the effective temperature (in K), the
  stellar mass (in units of solar mass); the initial helium content; the
  metallicity; the stellar radius (in units of solar radius); the logarithm of
  surface gravity (in cm s$^{-2}$).

\section{Results}\label{sec:results}

\begin{figure*}
\centering
\includegraphics[height=16cm,angle=-90]{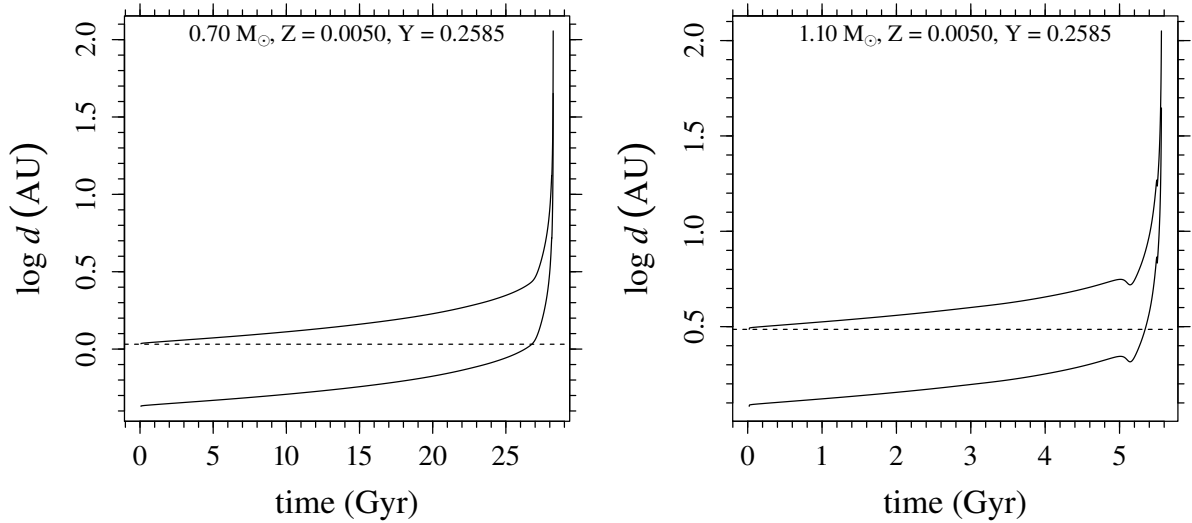}
\caption{Temporal evolution of the inner $d_{\rm i}$ and outer $d_{\rm o}$
  boundaries of
  the habitable zone for host 
  stars of $M$~=~0.70~$M_{\sun}$ (left panel) and  $M$~=~1.10~$M_{\sun}$
  (right panel) for $Z$~=~0.005, $Y$ = 0.2585 (i.e. $\Delta Y/\Delta Z = 2$), and $T_{\rm
    o}$~=~169~K.   
The HZ is the region between the two curves. The
  dashed lines mark the distance for which the HZ lasts longer.}
\label{fig:fascia}
\end{figure*}

In Fig.~\ref{fig:fascia} we display the temporal evolution of the inner and
outer boundaries -- $d_{\rm i}$ and $d_{\rm o}$ -- of the HZ,
derived from Eq.~(\ref{eq:d-HZ}) for stars with $M$~=~0.70 and 1.10~$M_{\sun}$
at $Z$~=~0.005 and $\Delta Y/\Delta Z$~=~2, assuming $T_{\rm 
  o}$~=~169~K for the equilibrium temperature of the outer boundary. 
Figure~\ref{fig:fascia-log} shows the same quantities but
with respect to the 
logarithm of time, allowing the inspection of the pre-main sequence phase.
The HZ is the region between the two boundaries. The figure shows that
excluding the early pre-main sequence phase, the HZ moves progressively
farther from the host star during its evolution as a consequence of the
continuously growing luminosity. The time scale of this displacement depends
on the evolutionary time scale of the host star.  Thus, it is long during the
main sequence phase and becomes much shorter after the central hydrogen
depletion.

The duration of habitability at distance $d$ is defined as the time interval for which
the inequality $d_{\rm i} \leq d \leq d_{\rm o}$ holds. If, for a given $d$, 
this inequality holds for multiple separated time intervals -- as in the left
panel of 
Fig.~\ref{fig:fascia-log} -- only the longest duration is considered.  The
dashed lines in the figures mark the distance $d_{\rm m}$ for which the HZ
has the maximum duration $t_{\rm m}$. Notice that $d_{\rm m}$ corresponds to the 
value of $d_{\rm o}$ when the host star is in its ZAMS stage.  

\onlfig{
\begin{figure*}
\centering
\includegraphics[height=16cm,angle=-90]{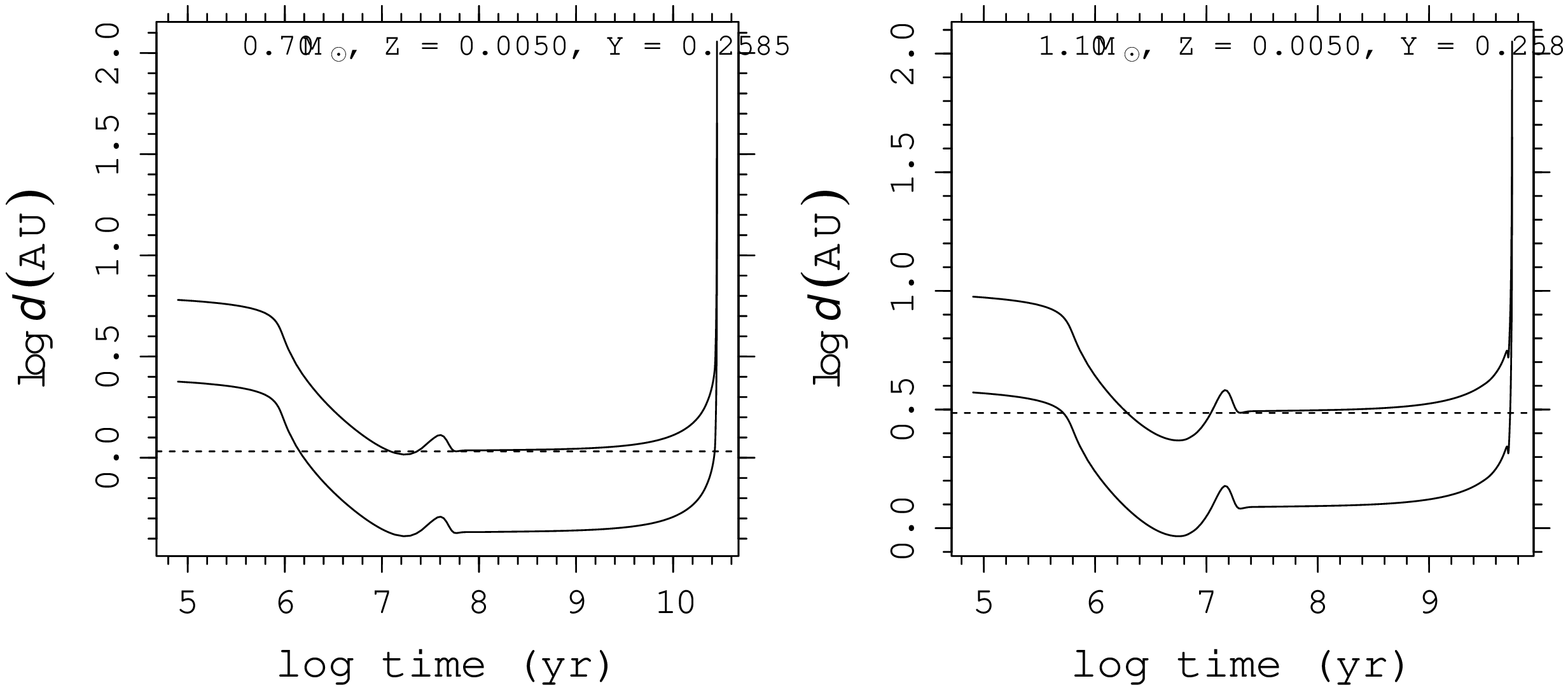}
\caption{As in Fig.~\ref{fig:fascia}, but for the logarithm of the time.}
\label{fig:fascia-log}
\end{figure*}
}

\begin{figure*}
\centering
\includegraphics[width=16cm,angle=-90]{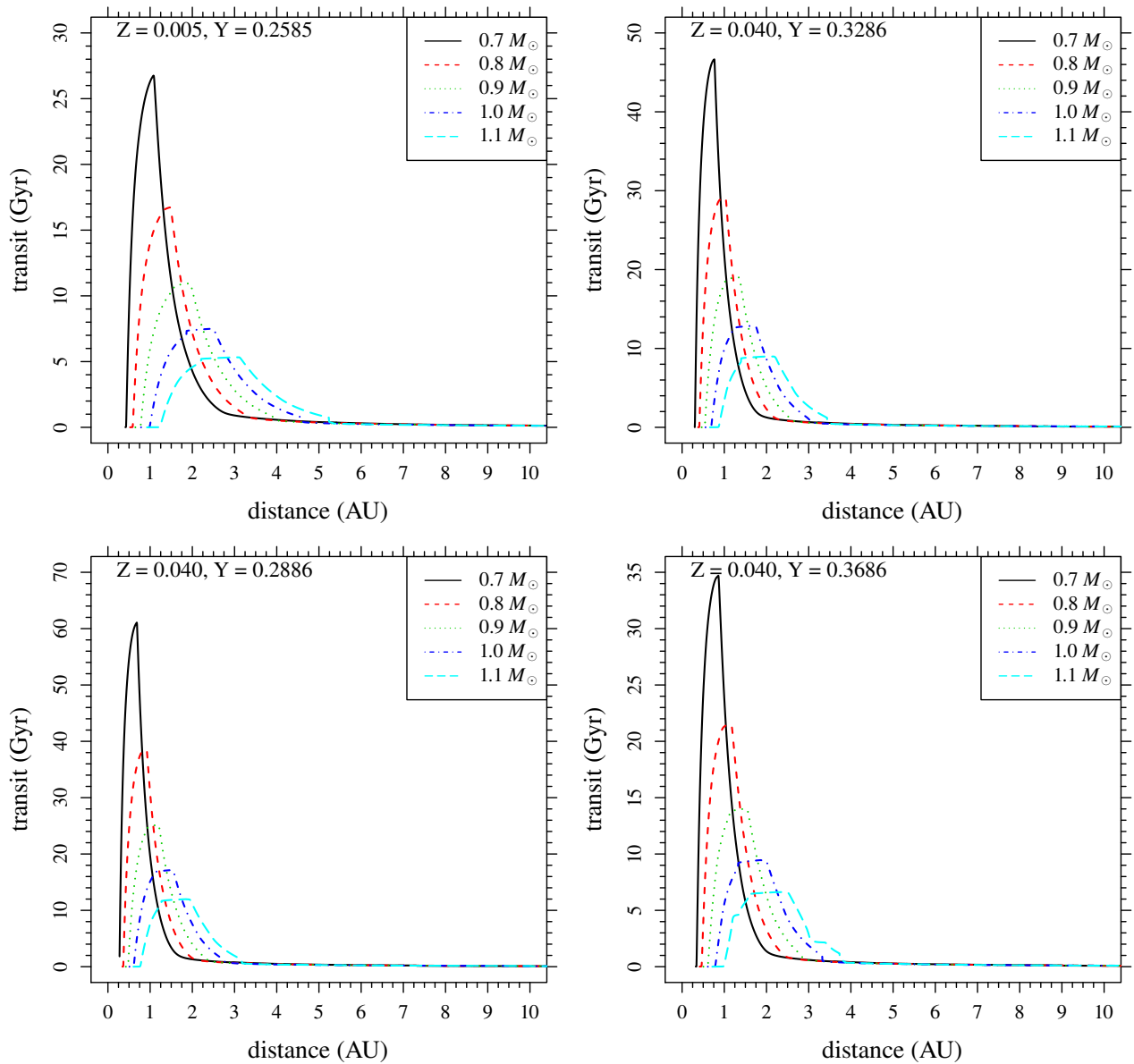}
\caption{(Top row): duration of habitability (transit)
as a function of the distance from the host star, for different masses $M$ 
and chemical compositions of the  host star for a fixed value of $\Delta Y/\Delta Z
= 2$ and two metallicities $Z$ = 0.005 and $Z$ = 0.040. (Bottom): same as the
top row, but for a fixed metallicity $Z$ = 0.040 and two values of $\Delta
Y/\Delta Z$, namely  1 (left
panel) and 3 (right
panel).}
\label{fig:transiti}
\end{figure*}

Figure~\ref{fig:transiti} shows the duration of the habitability, also known
as transit, as a function of the distance from the host star for different
stellar masses and chemical compositions. The two panels of the top row adopt
a fixed $\Delta Y/\Delta Z = 2$ but different metallicities, namely $Z =
0.005$ (left) and $Z = 0.04$ (right). The two panels of the bottom row have a
fixed metallicity $Z = 0.04$ but different $\Delta Y/\Delta Z$ values, namely
1 (left) and 3 (right). For a given chemical composition, the
more massive the host star, the farther the HZ, the larger $d_{\rm m}$,
and 
the shorter the transit duration $t_{\rm m}$.  This is the expected
consequence of the increasing brightness and decreasing lifetime of the host
star as the mass increases. The plateau around the longest duration for masses
higher than about 1.0~$M_{\sun}$ is due to the drop in the luminosity of these
stars after the central hydrogen depletion in the sub-giant branch (see
e.g. the right panel of Fig.~\ref{fig:fascia}, around 5.2~Gyr). From the
comparison of the two top panels it follows that a metallicity increase at
fixed $\Delta Y/\Delta Z$ leads to a decrease of $d_{\rm m}$ and to an increase
of the longest habitability duration $t_{\rm m}$.  This behaviour can be
easily understood by recalling that metal-rich stars evolve slower and fainter
than metal-poor ones of the same mass. The comparison of the two bottom
panels shows that a $\Delta Y/\Delta Z$ increase at fixed metallicity
$Z$ leads to an increase of $d_{\rm m}$ and a decrease of $t_{\rm m}$, that is,
the opposite effect of increasing $Z$. Again, this is the consequence of the
well-known dependence of stellar characteristics on the initial helium
abundance: 
at a given mass and $Z$, the higher $Y$ and the brighter and faster
the
evolution of a star. The steps that appear in the transit computed with high
initial helium abundance for $M$~=~1.1~$M_{\sun}$ are due to the development
of a convective core, which produces a step in the rate of 
the luminosity evolution with time.  Figure~\ref{fig:transiti-log} 
shows the same 
quantities, but for the logarithm of the transit time. The transformation
allows a better display of the transit during the red-giant branch
evolution of the host star.

The results presented here agree fairly well with those reported in
\citet{Danchi2013}. In the quoted paper, for an host star of 1.0~$M_{\sun}$
with 
metallicity $Z = 0.017$ and $Y = 0.26$, the distance corresponding to the
longest transit is around
1.7~AU, with a longest habitability duration  slightly higher than 10~Gyr. In
the present work, for a host star of 1.0~$M_{\sun}$, metallicity $Z = 0.015$,
and $Y = 0.263$ we found $d_{\rm m}$~=~1.9~AU, and $t_{\rm m}$~=~11.8~Gyr.
Other comparisons are less meaningful since the initial helium content for
different metallicities is chosen in a different way in the two
works.

\onlfig{
\begin{figure*}
\centering
\includegraphics[width=16cm,angle=-90]{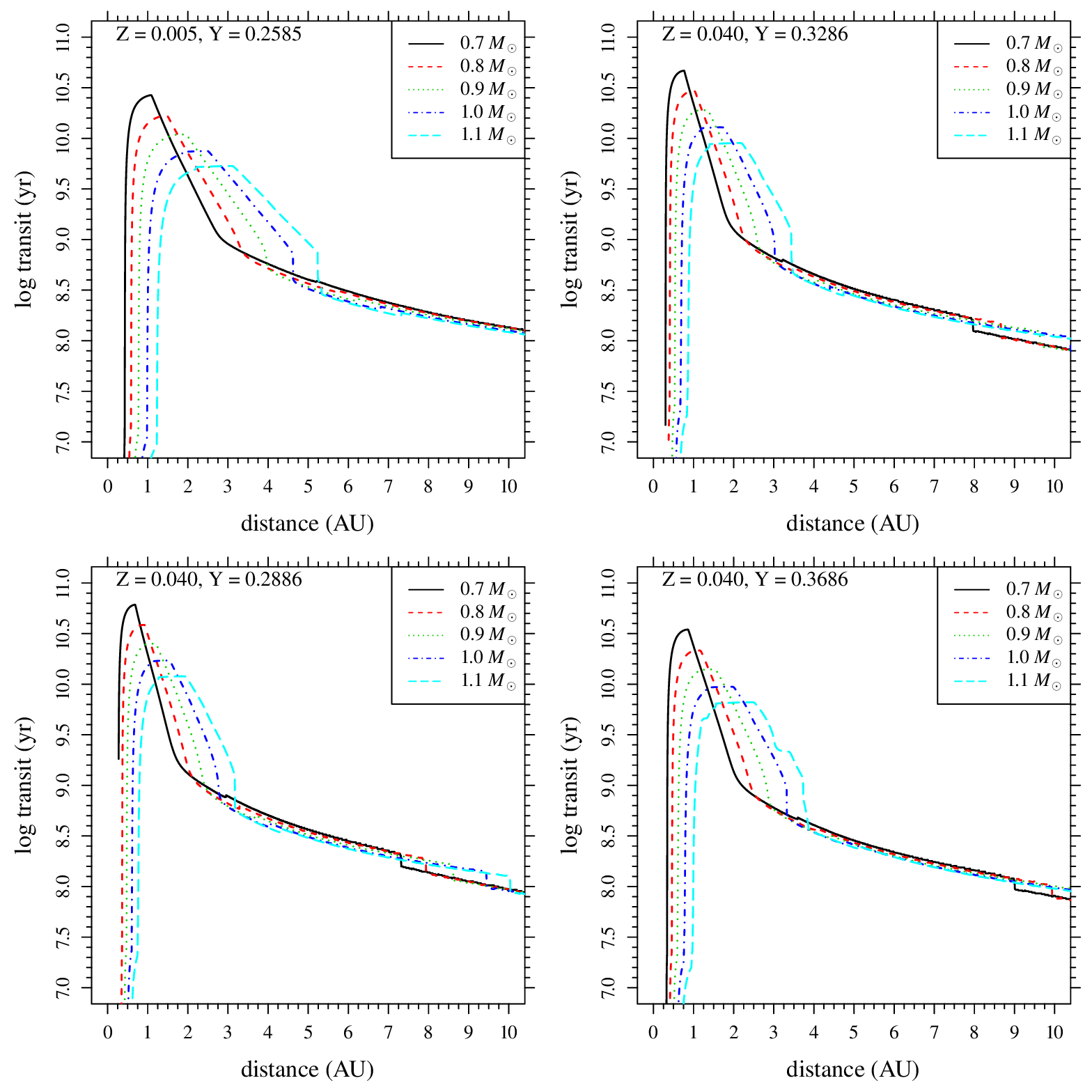}
\caption{As in Fig.~\ref{fig:transiti}, but for the logarithm of the transit
  time.} 
\label{fig:transiti-log}
\end{figure*}
}

We focused our analysis on some features of the HZ; in addition to
 $d_{\rm m}$ (in AU) and $t_{\rm m}$ (in Gyr) defined above, we examined 
the width $W$ (in AU) of the zone
for which the habitability lasts $t_{\rm m}/2$, and the integral $I$ (in AU
Gyr) of the transit function for transits longer than 4~Gyr.  These four
quantities allow characterising the transit function in the most
interesting region, where the transit time exceeds a few Gyr.

\begin{figure*}
\centering
\includegraphics[width=16cm,angle=-90]{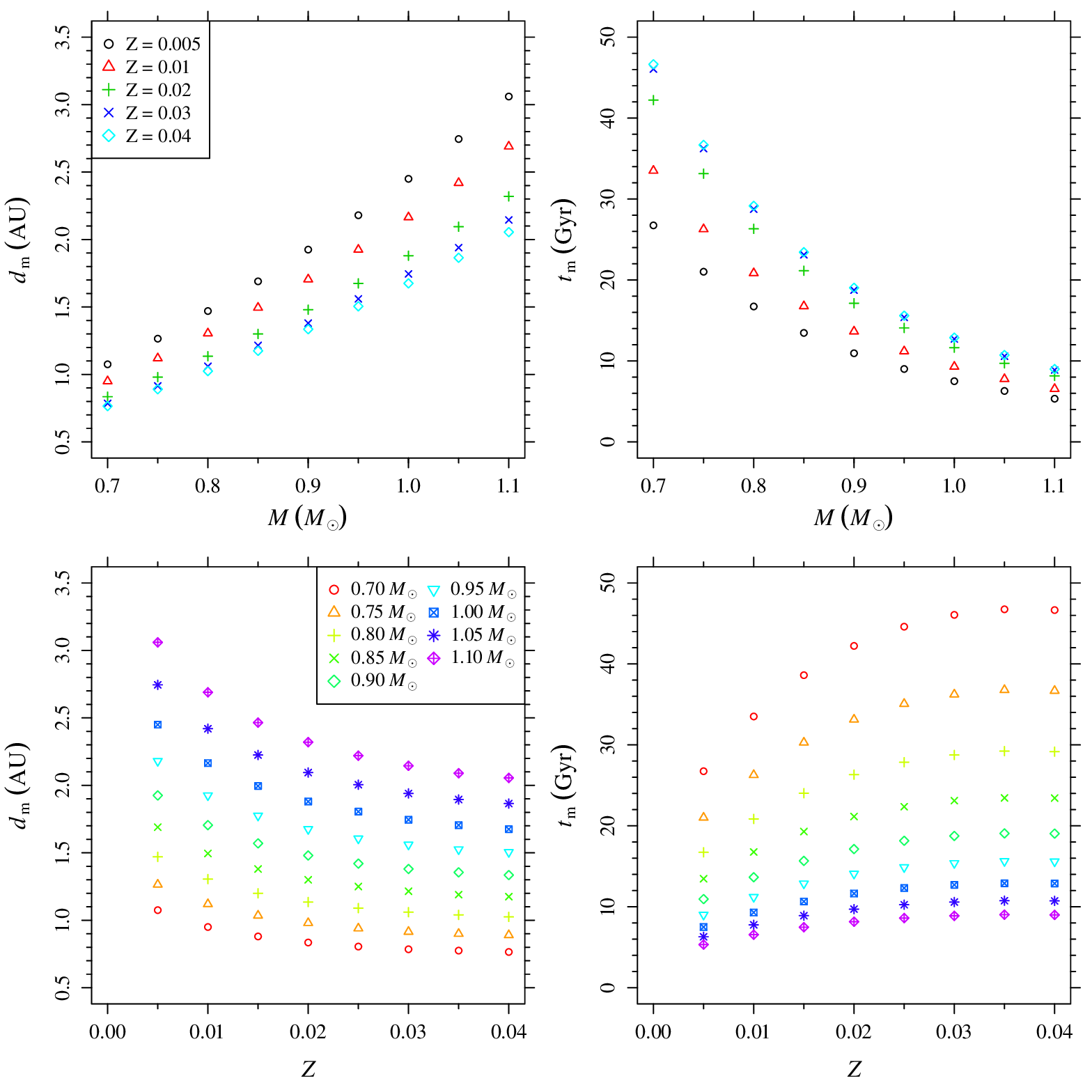}
\caption{(Top row): the distance $d_{\rm m}$ (in AU) for which the duration of
  the habitability is longest and the corresponding duration $t_{\rm m}$ (in
  Gyr) as a function of the mass of the host star computed for different
  metallicities $Z$.  (Bottom row): $d_{\rm m}$ (in AU) and $t_{\rm m}$ (in
  Gyr) as a function of the metallicity of the star computed for different
  masses $M$.  The value of initial helium abundance is derived assuming
  $\Delta Y/\Delta Z = 2$.}
\label{fig:XM}
\end{figure*}

\begin{figure*}
\centering
\includegraphics[width=16cm,angle=-90]{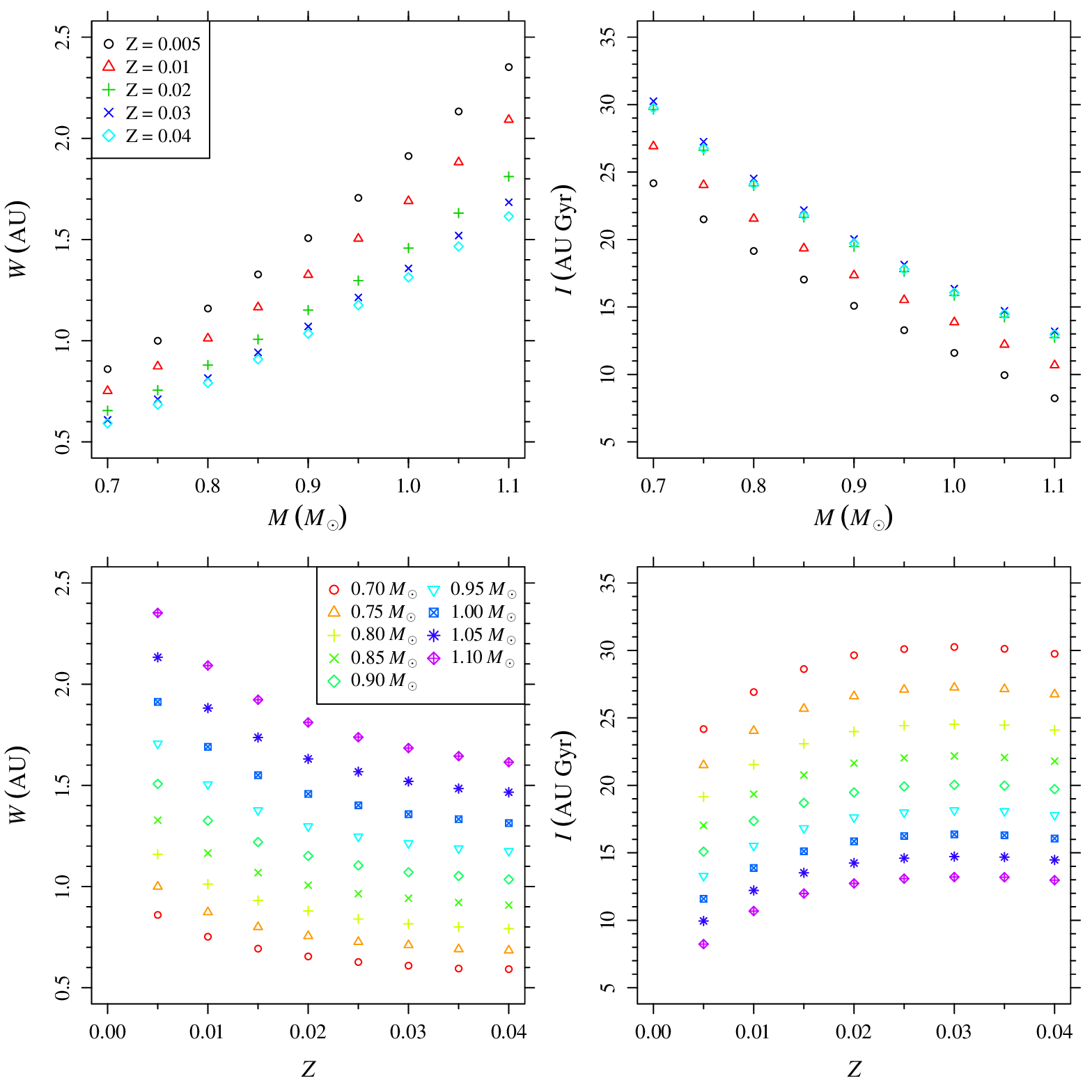}
\caption{As in Fig.~\ref{fig:XM}, but for $W$, i.e. the width (in AU) of the
  zone for which the habitability lasts $t_{\rm m}/2$, and $I$, i.e. the
  integral (in AU Gyr) of the transit function for transits longer than 4
  Gyr.}
\label{fig:WI}
\end{figure*}

Figures~\ref{fig:XM} and \ref{fig:WI} show the trend of the four chosen
features with respect to the mass and metallicity of the host star.
The distance $d_{\rm m}$ corresponding to the longest HZ duration
monotonically increases with the mass, 
while it decreases with $Z$ toward a plateau for $Z \geq 0.03$. The trend for
$W$ is the same as that for $d_{\rm m}$, while that for $t_{\rm m}$ is
reversed. For $I$, the effect of the metallicity is not monotonic: $I$ reaches
a maximum around $Z = 0.03$, with small variations in the
$Z$ range [0.02 - 0.04].  
This trend results from the combination of an increase with $Z$ of longest
time of habitability $t_{\rm m}$, which is steeper for a metallicity below about
$Z = 0.02$, with the trend of $d_{\rm m}$, which changes more smoothly with
metallicity. 

A high value of $I$ is preferable in the observational
target selection, since it is related to two desirable characteristics.
The larger $I$, the larger either the
extension of the 4~Gyr CHZ -- implying higher chances that an exoplanet actually
lies inside it --  or the longer the habitability duration inside
the CHZ -- implying higher chances that an exoplanet in the 4~Gyr CHZ is indeed
in a region that is continuously habitable for more than 4~Gyr.   
Since the curves in the different panels of Fig.~\ref{fig:XM} and \ref{fig:WI}
are not 
parallel, there is an effect due to the interaction between mass and
metallicity, which means that the effect of changing the former quantity also depends on the
value of the latter.

The quantities discussed above are of relevant interest in
  the framework of the analysis of the impact of the stellar evolution on the
  HZ. However, to plan of a survey for life 
  signature in exoplanet atmospheres, the most interesting feature  is the
  CHZ boundaries position.  
  
\begin{figure*}
\centering
\includegraphics[width=8cm,angle=-90]{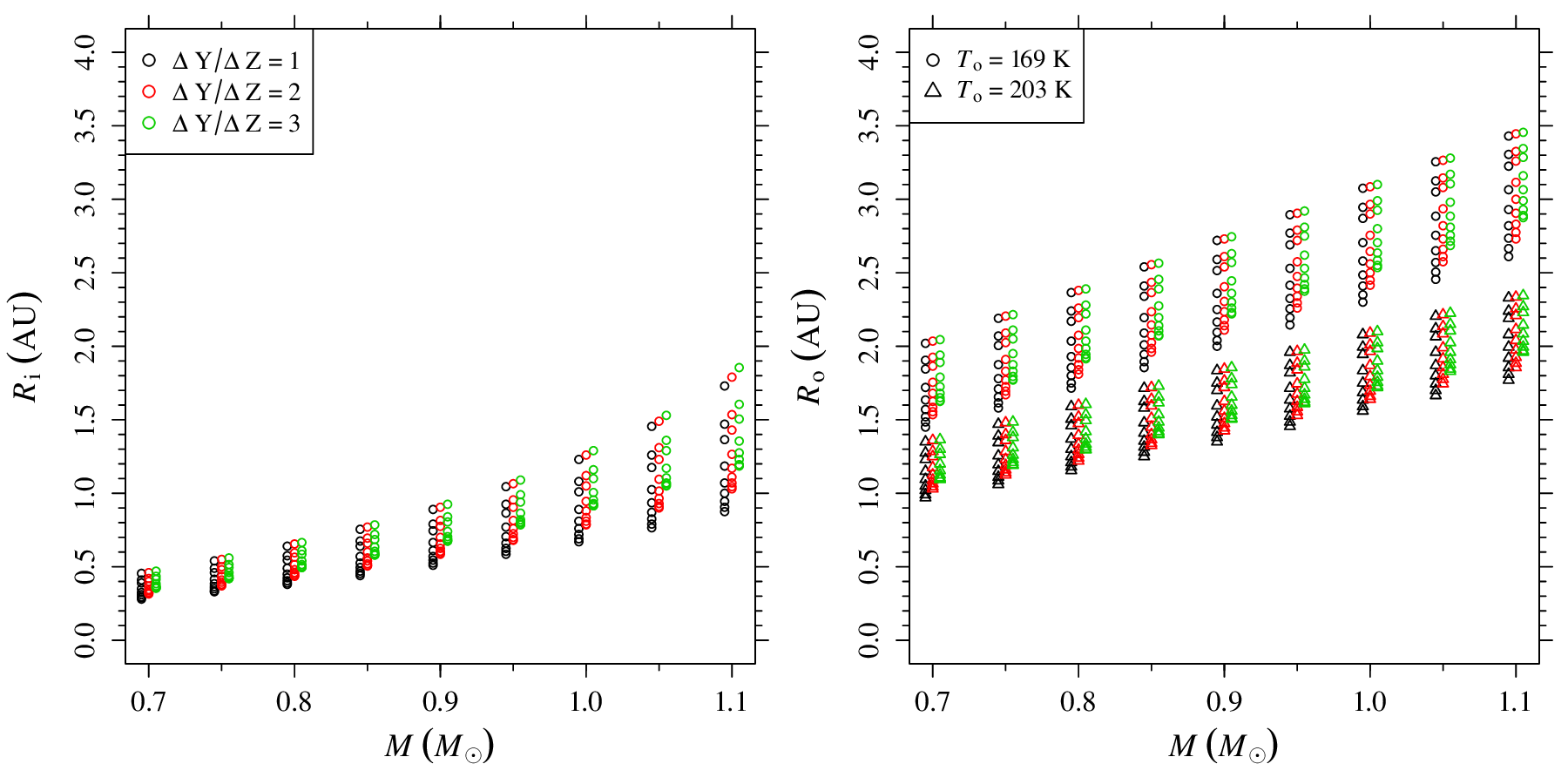}
\caption{(Left): position (in AU) of the 4~Gyr CHZ inner boundary as a
  function of the host star mass for the labelled $\Delta Y/\Delta Z$
  values. For each set of mass and $\Delta Y/\Delta Z$, the metallicity $Z$
  runs from 0.04 at the upper point to 0.005 at the lower one. The initial
  helium content is obtained from $Z$ and $\Delta Y/\Delta Z$ as in Eq.~(\ref{eq:YZ}).
To show the effect of the initial helium
  content, the abscissa of
  models with high (low) $\Delta Y/\Delta Z$ are shifted by adding
  (subtracting) 0.005~$M_{\sun}$. (Right): same as the left panel, 
  but for the outer boundary. The circles correspond to the computations with
  $T_{\rm o}$~=~169~K, while the triangles correspond to those with $T_{\rm
    o}$~=~203~K.} 
\label{fig:chz-bound}
\end{figure*}

Figure~\ref{fig:chz-bound} shows the position of the inner ($R_{\rm i}$, left
panel) and 
outer ($R_{\rm o}$, right panel) boundaries of the 4~Gyr CHZ as a function of
the host star 
mass for three $\Delta Y/\Delta Z$ values, namely 1, 2, and 3.  
This allows exploring the effect of changing the initial helium
  abundance following Eq.~(\ref{eq:YZ}).
To
improve the figure readability the abscissa of stars with high (low)
helium-to-metal 
enrichment ratio is shifted by adding (subtracting)
0.005~$M_{\sun}$.  
As expected, $R_{\rm i}$ and $R_{\rm o}$ increase with mass and decrease with
metallicity.
The temperature $T_{\rm o}$ of the outer boundary has no
influence on the determination of the inner boundary.  
The spread of $R_{\rm i}$ and $R_{\rm o}$ with metallicity increases when mass
and $\Delta Y/\Delta Z$ increase.
This suggests that interaction of metallicity, mass and helium content are
important in determining the boundaries. 
The effect of a change of $\Delta
Y/\Delta Z$ for $M$~=~1.1~$M_{\sun}$ can be as large as
0.32~AU for $Z = 0.04$, while it decreases to a maximum of 0.12~AU for  $Z =
0.005$.  
In the right panel of the figure we also show the effect of the change of
$T_{\rm o}$ on the outer boundary. In this case, the effect of the change of
the initial helium content is as large as 0.26~AU for $Z = 0.04$ and 
$M$~=~1.1~$M_{\sun}$ in the case of $T_{\rm o}$~=~169~K, and it decreases to a
maximum of 0.19~AU for $Z = 0.04$ and $T_{\rm o}$~=~203~K. For the lowest
analysed metallicity, $Z = 0.005$, these values drop to 0.025~AU and 0.015~AU.

\begin{figure*}
\centering
\includegraphics[width=8cm,angle=-90]{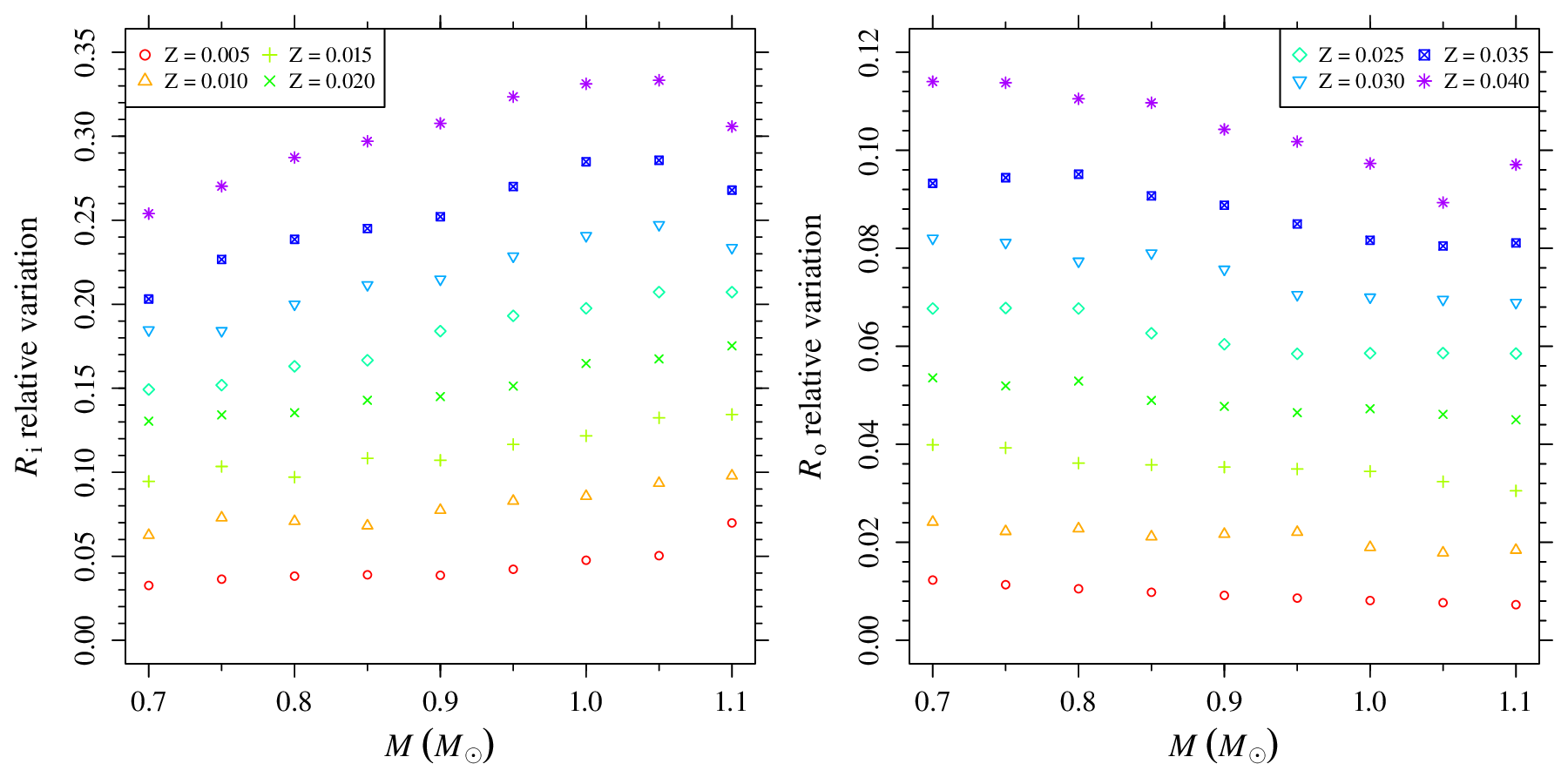}
\caption{(Left): relative variation for a change from $\Delta Y/\Delta Z = 3$
  to $\Delta Y/\Delta Z = 1$ of the 4~Gyr CHZ inner boundary as a
  function of the host star mass for the labelled $Z$
  values. (Right): same as the left panel,
  but for the outer boundary. The adopted outer boundary temperature is
  $T_{\rm o}$~=~169~K.}
\label{fig:chz-bound-relerr}
\end{figure*}

The effect of varying the initial helium content is also displayed in
Fig.~\ref{fig:chz-bound-relerr}, which shows the relative variation of the
inner and 
outer boundaries of the CHZ for a change on the helium-to-metal enrichment
from $\Delta Y/\Delta Z = 3$ to 
$\Delta Y/\Delta Z = 1$. For $R_{\rm i}$ (left panel) the relative
variation 
ranges from about 5\% for $Z = 0.005$ to about 30\% for $Z$~=~0.04, and
it generally increases with the mass of the star. The
decrease of variation that occurs for models of 1.1~$M_{\sun}$ at high
metallicities are due to the development of a convective core for the
helium-rich models, which causes a step in the transit curve (see the bottom
row in Fig.~\ref{fig:transiti}).
For $R_{\rm o}$ (left panel in Fig.~\ref{fig:chz-bound-relerr}), assuming
$T_{\rm o}$~=~169~K,  the relative
variation 
ranges from about 3\% for $Z = 0.005$ to about 10\% for $Z = 0.04$, and
it generally mildly decreases with the mass of the star. As in the case
discussed above, the exception are the models of 1.1~$M_{\sun}$ at
high metallicity.

The comparison of the result obtained in this work for a 1.0~$M_{\sun}$ model
with the corresponding one presented in
\citet{Danchi2013} -- who adopted a similar approach -- requires some
ad hoc calculations since in the quoted paper the boundaries of CHZ was
computed for durations of 1 
and 3~Gyr. The CHZ boundaries 
for an host star of 1.0~$M_{\sun}$, $Z = 0.017$, and $Y = 0.26$
were reported to be [0.8 - 3.3]~AU, and [0.9 - 2.8]~AU for a duration of 1 and
3~Gyr, respectively. 
The corresponding values computed ad-hoc 
for the model of 
1.0~$M_{\sun}$, $Z = 0.015$,
and $Y = 0.263$ are [0.84 - 3.56]~AU and [0.90 - 2.94]~AU, respectively.

\section{Analytical models}\label{sec:modelli}

The dataset of computed stellar models allows a statistical analysis of
the dependence of the previously discussed features on the host stellar
characteristics.  
This analysis has three purposes. First, it allows distinguishing the
  contribution of the various input to the temporal variation of the HZ and
  the CHZ positions. Second, it provides an accurate and precise analytical 
  model that can be safely used to evaluate the required characteristics
  without the need of  
  time-consuming stellar model computations. As already discussed
   in Sect.~\ref{sec:methods},  more than on
actual estimates resulting from
    the models, we are interested in showing the degree of
    precision that can be reached -- given a HZ defining scenario -- by
    analytical relations. 
Third, these accurate  models allow an uncertainty analysis to be performed
(see Sect.~\ref{sec:sensitivity}) without the need to compute
the huge number of detailed stellar models that would otherwise have been 
required.

The analysis was conducted by adapting linear models on the
data, assuming $d_{\rm m}$, $t_{\rm m}$, $W$, $R_{\rm i}$, and $R_{\rm o}$ as
dependent variables and 
the mass of the host star $M$, the metallicity $Z$, and the initial helium
abundance $Y$ as independent variables (or covariates). The temperature of the
outer boundary $T_{\rm o}$ is also inserted in the statistical models.
Including this quantity in the analytical models allows the description
of the HZ characteristics for a wide range of assumption on $T_{\rm
  o}$. This is especially useful since the determination of the outer
boundary temperature strongly depends on the cloud coverage, which is not
accurately described in the 1D climate models.

The building of the statistical models requires several checks.  We started
with simple 
relations including linearly the covariates, but allowing for interaction
between chemical inputs ($Z$ and $Y$), the mass of the star, and the boundary
temperature. Then we checked whether the model was able to describe all
significant trends in the data without over-fitting them \citep[see
  e.g.][]{linmodR}.  The first requirement was tackled by the analysis of the
standardised residuals, to check that the entire information of the
data is extracted by the model. The plots of the standardised residuals
versus the covariates were used to infer the need to include quadratic or
cubic terms or high-order interactions. 
As a result we obtained that although quadratic models
  are able 
to explain the bulk of data variability, they are statistically
unacceptable 
since clear trends remain in the residuals. Cubic terms
and more complex interactions are therefore needed in the models.    
 The problem of possible over-fitting
required using the stepwise regression \citep{venables2002modern}
technique, which allows one to evaluate the performance of the multivariate
model (balancing the goodness-of-fit and the number of covariates in the
model) and of the models nested in this one (i.e. models without some of the
covariates).  To perform the stepwise model selection we employed the Bayesian
information criterion (BIC):
\begin{equation}
{\rm BIC} = n \log \frac{d^2_E}{n} + p \; \ln n \;,
\end{equation}
which balances the number of covariates $p$ included in the model and its
performance in the data description, measured by the error deviance $d^2_E$
($n$ is the number of points in the model). Among the models explored by the
stepwise technique we selected that with the lowest BIC value as the
best one.
The models were fitted to the data with a least-squares method using the
software R 3.0.2 \citep{R}.

An interactive web code and a C program to obtain the HZ and CHZ
characteristics described by 
the analytical models presented in this section are available
on-line\footnote{\url{http://astro.df.unipi.it/stellar-models/HZ/}}. The
  web interface also computes an interpolated stellar track -- from pre-main
  sequence to the helium flash at the red-giant branch tip -- for the model of
  mass, metallicity, and initial helium content required by the user.
The file provide as a
  function of time (in Gyr) the logarithm of stellar luminosity (in units
  of solar luminosity); the logarithm of the effective temperature (in K), the
  stellar mass (in units of solar mass); the initial helium content; the
  metallicity; the stellar radius (in units of solar radius); the logarithm of
  surface gravity (in cm s$^{-2}$). The values can be useful for computing
  synthetic spectra of the incoming radiation to be coupled to climate models.

To concisely describe the analytical models, in this section we use the
operator $*$, 
defined as $A * B \equiv A + B + A \cdot B$, and we excluded
the regression coefficients in the models. The full forms of the models are
reported in Appendix~\ref{sec:append-models}.

\begin{figure*}
\centering
\includegraphics[width=6cm,angle=-90]{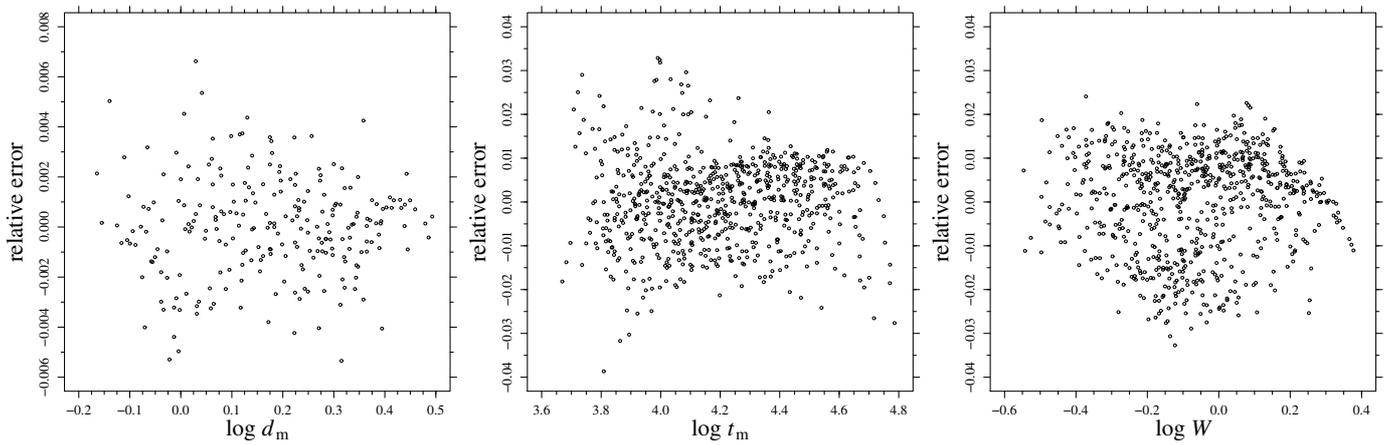}
\caption{(Left): relative errors (positive values correspond to overestimated
  values) on the distance $d_{\rm m}$ at which the HZ 
  has the longest 
duration, as estimated from the model in Eq.~(\ref{eq:mod-xm}). (Middle):
relative errors on the longest duration $t_{\rm m}$, as estimated from the
model in Eq.~(\ref{eq:mod-ym}).  (Right): relative errors on width at half
maximum 
$W$, as estimated from the 
model in Eq.~(\ref{eq:mod-W}).}
\label{fig:relative-error-xyW}
\end{figure*}

The analytical model for the distance $d_{\rm m}$ at which the HZ has the longest
duration is
\begin{equation}
\log d_{\rm m} ({\rm AU}) = K_1 + (M * Z_{\rm l} + M^2 * Z_{\rm l}^2 + M^3 * Z_{\rm
  l}^3) * Y \; ,
\label{eq:mod-xm}
\end{equation}
where $Z_{\rm l} = \log Z$, and $K_1$ is a scaling factor that takes into
account different choices for the values of albedo $A$ and $T_{\rm o}$: $K_1
= \log \left(\sqrt{(1-A)/0.7} \; /T_{169}^2 \right)$, $T_{169} = T_{\rm
  o}/(169 \; {\rm K})$.  The terms that are selected in the stepwise procedure
described above are reported in Tab.~\ref{table:mod-xm}. 
The first two columns of the table report the
least-squares estimates of the regression coefficients and their errors; the
third column reports the $t$-statistic for the tests of the statistical
significance of the covariates; the fourth column reports the
$p$-values of these tests. 
The left panel of
Fig.~\ref{fig:relative-error-xyW} shows the relative error on
the position of the maximum, i.e.  $\frac{\hat d_{\rm m}-d_{\rm m}}{d_{\rm
    m}}$ where $\hat d_{\rm m}$ is the position estimated from the linear
model, as a function of $\log d_{\rm m}$. The relative errors are seldom
greater -- in modulus -- than 0.4\%, implying a good accuracy of the estimates
provided by Eq.~(\ref{eq:mod-xm}). 

The analytical model for the longest duration $t_{\rm m}$ (in Gyr) of the HZ is
\begin{equation}
\log t_{\rm m} ({\rm Gyr}) = (\frac{1}{M} + M_{\rm l} * Z_{\rm l} * Y +  M_{\rm l}^2 *
Z_{\rm l}^2) * (T_{169} + T_{169}^2) \; ,
\label{eq:mod-ym}
\end{equation}
where $M_{\rm l} = \log M/M_{\odot}$. In this model the
dependence on $T_{169}$ must be explicitly included since there is no simple
geometrical scaling as in the case of $d_{\rm m}$.  The terms
selected in the stepwise procedure described above are reported in
Tab.~\ref{table:mod-ym}. The central panel of
Fig.~\ref{fig:relative-error-xyW} show the relative error on $t_{\rm m}$ as
resulting from the estimates of Eq.~(\ref{eq:mod-ym}). In this case, the model
relative errors are greater than the ones for $d_{\rm m}$, but are seldom
greater than 2\% in modulus.

The analytical model for $W$ (in AU) of the HZ is
\begin{equation}
\log W ({\rm AU}) = K_2 + (M * Z_{\rm l} + M^2 * Z_{\rm l}^2 + Z_{\rm l}^3)* Y
* T_{169} \; ,
\label{eq:mod-W}
\end{equation}
where the scaling $K_2 =
\log \sqrt{(1-A)/0.7}$ accounts for possible different choices of the albedo
$A$. 
 The terms selected in the stepwise procedure are
reported in Tab.~\ref{table:mod-W}. The right panel of
Fig.~\ref{fig:relative-error-xyW} shows the relative error on $W$ as
resulting from the estimates of Eq.~(\ref{eq:mod-W}). The dispersion of the
relative errors is almost of the same magnitude of the one for $t_{\rm m}$.

\begin{figure*}
\centering
\includegraphics[width=8cm,angle=-90]{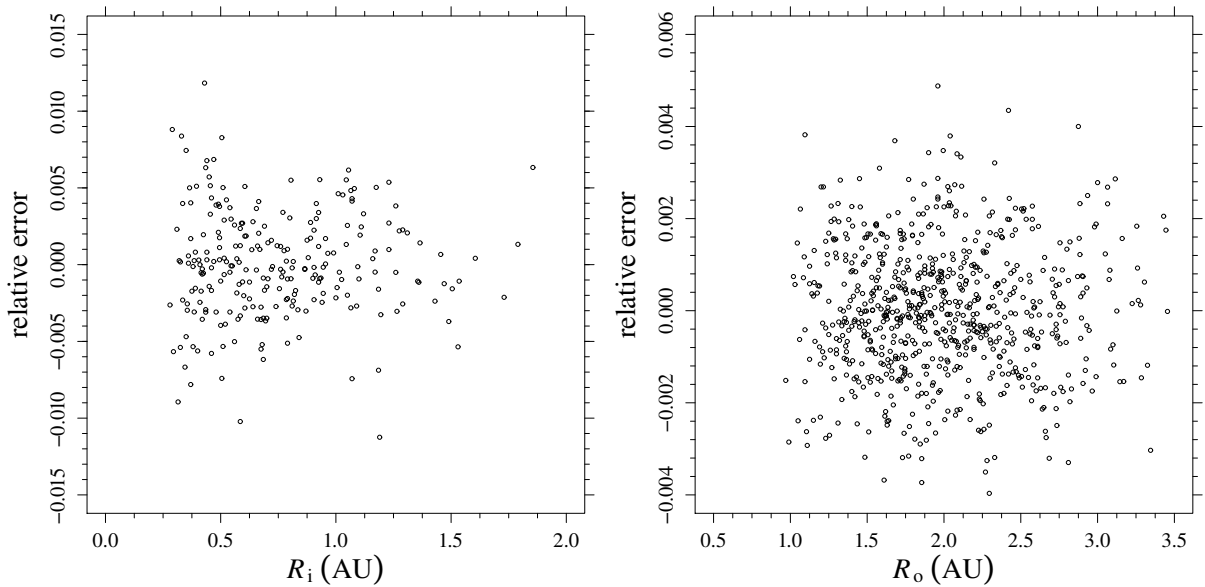}
\caption{(Left): relative errors on the distance $R_{\rm i}$ of the  inner 
boundary of the 4~Gyr CHZ,
as estimated from the analytical model in Eq.~(\ref{eq:mod-rl}). (Right): 
as in the left panel, but for the CHZ outer boundary $R_{\rm o}$,  
as estimated from the analytical model in Eq.~(\ref{eq:mod-ru}).}
\label{fig:chz-residui}
\end{figure*}

Regarding the 4~Gyr CHZ, 
the analytical model for the inner boundary $R_{\rm i}$ is
\begin{equation}
\log R_{\rm i} ({\rm AU}) = K_2 + (M * Z_{\rm l} + M^2 * Z_{\rm l}^2 + M^3 * Z_{\rm
  l}^3) * Y \; .
\label{eq:mod-rl}
\end{equation}
The terms that are selected in the stepwise procedure are
reported in Tab.~\ref{table:mod-rl}.
The left panel of
Fig.~\ref{fig:chz-residui} shows the relative error in $R_{\rm i}$ as
resulting from the estimates of Eq.~(\ref{eq:mod-rl}). The relative error is
generally lower than 0.5\%, so the model description 
of the CHZ inner boundary trend is fairly accurate.

For the outer boundary $R_{\rm o}$, the dependence on the outer
temperature $T_{169}$ is included in the model:
\begin{equation}
\log R_{\rm o} ({\rm AU}) = K_2 + (M * Z_{\rm l} + M^2 * Z_{\rm l}^2 + M^3 * Z_{\rm
  l}^3) * Y * T_{169} + T_{169}^2 \; .
\label{eq:mod-ru}
\end{equation}
The terms selected in the stepwise procedure are
reported in Tab.~\ref{table:mod-ru}.
The relative error on $R_{\rm o}$, as
resulting from the estimates of Eq.~(\ref{eq:mod-rl}), is shown in the right
panel of Fig.~\ref{fig:chz-residui}. The relative error is
generally lower than 0.3\%, so the outer boundary is described with higher
accuracy than the inner one.

\begin{table}[ht]
\centering
\caption{Regression terms of the model of $d_{\rm m}$ in Eq.~(\ref{eq:mod-xm})
  which are 
  selected in the stepwise procedure according to the BIC index.} 
\label{table:mod-xm}
\begin{tabular}{rrrrr}
  \hline\hline
 & Estimate & Std. error & $t$ value & $p$ value \\ 
  \hline
(Intercept) & -1.2410 & 0.1447 & -8.58 & 1.53 $\times 10^{-15}$ \\ 
  $M$ & 1.5820 & 0.2579 & 6.13 & 3.75 $\times 10^{-9}$ \\ 
  $M^2$ & -1.2290 & 0.1614 & -7.62 & 6.90 $\times 10^{-13}$ \\ 
  $M^3$ & 0.4822 & 0.0555 & 8.69 & 7.37 $\times 10^{-16}$ \\ 
  $Z_{\rm l}$ & 1.0791 & 0.1279 & 8.44 & 3.79 $\times 10^{-15}$ \\ 
  $Z_{\rm l}^2$ & 0.5470 & 0.0382 & 14.32 & $< 2.0 \times 10^{-16}$ \\ 
  $Z_{\rm l}^3$ & 0.0952 & 0.0055 & 17.37 & $< 2.0 \times 10^{-16}$ \\ 
  $Y$ & 1.0832 & 0.0554 & 19.56 & $< 2.0 \times 10^{-16}$ \\ 
  $M \cdot Z_{\rm l}$ & -1.0129 & 0.1373 & -7.38 & 2.99 $\times 10^{-12}$ \\ 
  $Y \cdot  Z_{\rm l}^3$ & 0.0051 & 0.0049 & 1.05 & 2.94 $\times 10^{-1}$ \\ 
  $M^2 \cdot Z_{\rm l}^2$ & -0.2799 & 0.0409 & -6.85 & 6.96 $\times 10^{-11}$ \\ 
  $Y \cdot M^2$ & 0.8673 & 0.1724 & 5.03 & 9.91 $\times 10^{-7}$ \\ 
  $Y \cdot M^3$ & -0.8351 & 0.1178 & -7.09 & 1.66 $\times 10^{-11}$ \\ 
  $M^3 \cdot Z_{\rm l}^3$ & -0.0285 & 0.0056 & -5.07 & 8.27 $\times 10^{-7}$ \\ 
  $Y \cdot M^3 \cdot Z_{\rm l}^3$ & -0.0232 & 0.0057 & -4.10 & 5.76 $\times 10^{-5}$ \\ 
   \hline
\end{tabular}
\tablefoot{
In the first two columns we list the least-squares estimates of the regression
coefficients and their errors;   
third column: $t$-statistic for the tests of the statistical 
significance of the covariates; fourth column: $p$-values of the tests.}
\end{table}

\begin{table}[ht]
\centering
\caption{As in Table \ref{table:mod-xm} but for $t_{\rm m}$ in
  Eq.~(\ref{eq:mod-ym}).} 
\label{table:mod-ym}
\begin{tabular}{rrrrr}
  \hline\hline
 & Estimate & Std. error & $t$ value & $p$ value \\ 
  \hline
(Intercept) & -37.2982 & 3.7710 & -9.10 & $< 2.0 \times 10^{-16}$ \\ 
  $M_{\rm l}$ & 109.9736 & 9.9234 & 11.08 & $< 2.0 \times 10^{-16}$ \\ 
  $1/M$ & 40.7217 & 3.7594 & 10.83 & $< 2.0 \times 10^{-16}$ \\ 
  $Z_{\rm l}$ & 1.5098 & 0.0617 & 24.48 & $< 2.0 \times 10^{-16}$ \\ 
  $Z_{\rm l}^2$ & 0.0665 & 0.0403 & 1.65 & 9.98 $\times 10^{-2}$ \\ 
  $Y$ & -4.7661 & 0.0970 & -49.13 & $< 2.0 \times 10^{-16}$ \\ 
  $Y \cdot M_{\rm l}$ & -3.1963 & 0.5521 & -5.79 & 1.06 $\times 10^{-8}$ \\ 
  $Y \cdot Z_{\rm l}$ & -0.8552 & 0.0495 & -17.26 & $< 2.0 \times 10^{-16}$ \\ 
  $T_{169}$ & 77.5873 & 6.8519 & 11.32 & $< 2.0 \times 10^{-16}$ \\ 
  $T_{169}^2$ & -36.3864 & 3.0971 & -11.75 & $< 2.0 \times 10^{-16}$ \\ 
  $M_{\rm l} \cdot T_{169}$ & -212.5445 & 18.0314 & -11.79 & $< 2.0 \times 10^{-16}$ \\ 
  $M_{\rm l} \cdot T_{169}^2$ & 98.7415 & 8.1502 & 12.12 & $< 2.0 \times 10^{-16}$ \\ 
  $1/M \cdot T_{169}$ & -76.4845 & 6.8319 & -11.20 & $< 2.0 \times 10^{-16}$ \\ 
  $1/M \cdot T_{169}^2$ & 35.5695 & 3.0876 & 11.52 & $< 2.0 \times 10^{-16}$ \\ 
  $Z_{\rm l} \cdot T_{169}^2$ & -0.1967 & 0.0467 & -4.21 & 2.91 $\times 10^{-5}$ \\  
  $Z_{\rm l}^2 \cdot T_{169}$ & 0.2764 & 0.0677 & 4.08 & 4.99 $\times 10^{-5}$ \\ 
  $Z_{\rm l}^2 \cdot T_{169}^2$ & -0.1871 & 0.0330 & -5.66 & 2.15 $\times 10^{-8}$ \\ 
  $Y \cdot T_{169}^2$ & 0.3196 & 0.0497 & 6.43 & 2.40 $\times 10^{-10}$ \\ 
  $Y \cdot M_{\rm l} \cdot T_{169}^2$ & 2.1374 & 0.4493 & 4.76 & 2.39 $\times 10^{-6}$ \\ 
   \hline
\end{tabular}
\end{table}

\begin{table}[ht]
\centering
\caption{As in Table \ref{table:mod-xm} but for $W$ in
  Eq.~(\ref{eq:mod-W}).} 
\label{table:mod-W}
\begin{tabular}{rrrrr}
 \hline\hline
 & Estimate & Std. error & $t$ value & $p$ value \\ 
  \hline
(Intercept) & 0.6789 & 0.2210 & 3.07 & 2.21 $\times 10^{-3}$ \\ 
  $M$ & -0.9764 & 0.3488 & -2.80 & 5.26 $\times 10^{-3}$ \\ 
  $M^2$ & 1.1740 & 0.1661 & 7.07 & 3.70 $\times 10^{-12}$ \\ 
  $Z_{\rm l}$ & 0.3371 & 0.1032 & 3.27 & 1.14 $\times 10^{-3}$ \\ 
  $Z_{\rm l}^2$ & 0.4138 & 0.0564 & 7.34 & 5.92 $\times 10^{-13}$ \\ 
  $Z_{\rm l}^3$ & 0.0942 & 0.0102 & 9.22 & $< 2.0 \times 10^{-16}$ \\ 
  $Y$ & -0.2347 & 0.5653 & -0.42 & 6.78 $\times 10^{-1}$ \\ 
  $Y \cdot M$ & 2.1227 & 0.6155 & 3.45 & 5.96 $\times 10^{-4}$ \\ 
  $T_{169}$ & -2.6487 & 0.1924 & -13.77 & $< 2.0 \times 10^{-16}$ \\   
  $M \cdot T_{169}$ & 3.0859 & 0.3165 & 9.75 & $< 2.0 \times 10^{-16}$ \\ 
  $M^2 \cdot T_{169}$ & -1.7081 & 0.1507 & -11.33 & $< 2.0 \times 10^{-16}$ \\ 
  $Z_{\rm l}^3 \cdot T_{169}$ & -0.0065 & 0.0010 & -6.32 & 4.65 $\times 10^{-10}$\\
  $Y \cdot T_{169}$ & 1.6171 & 0.5130 & 3.15 & 1.69 $\times 10^{-3}$ \\ 
  $Y \cdot M \cdot T_{169}$  & -2.2411 & 0.5586 & -4.01 & 6.65 $\times 10^{-5}$ \\
   \hline
\end{tabular}
\end{table}

\begin{table}[ht]
\centering
\caption{As in Table \ref{table:mod-xm} but for $R_{\rm i}$ in
  Eq.~(\ref{eq:mod-rl}).} 
\label{table:mod-rl}
\begin{tabular}{rrrrr}
  \hline\hline
& Estimate & Std. error & $t$ value & $p$ value \\ 
  \hline
(Intercept) & -1.1957 & 1.2316 & -0.97 & 3.33 $\times 10^{-1}$ \\ 
  $M$ & -0.2712 & 2.4903 & -0.11 & 9.13 $\times 10^{-1}$ \\ 
  $M^2$ & -5.2040 & 2.0755 & -2.51 & 1.29 $\times 10^{-2}$ \\ 
  $M^3$ & 2.7739 & 0.7075 & 3.92 & 1.18 $\times 10^{-4}$ \\ 
  $Z_{\rm l}$ & 0.7318 & 1.1173 & 0.65 & 5.13 $\times 10^{-1}$ \\ 
  $Z_{\rm l}^2$ & -0.9036 & 0.4969 & -1.82 & 7.03 $\times 10^{-2}$ \\ 
  $Z_{\rm l}^3$ & -0.2442 & 0.0916 & -2.67 & 8.22 $\times 10^{-3}$ \\ 
  $Y$ & 4.8948 & 3.6152 & 1.35 & 1.77 $\times 10^{-1}$ \\ 
  $M \cdot Z_{\rm l}$ & -5.4295 & 0.8501 & -6.39 & 9.66 $\times 10^{-10}$ \\ 
  $M^2 \cdot Z_{\rm l}^2$ & -1.3059 & 0.1821 & -7.17 & 1.08 $\times 10^{-11}$ \\ 
  $M^3 \cdot Z_{\rm l}^3$ & -0.1098 & 0.0136 & -8.08 & 3.94 $\times 10^{-14}$ \\ 
  $Y \cdot M$ & -3.6520 & 7.2591 & -0.50 & 6.15 $\times 10^{-1}$ \\ 
  $Y \cdot M^2$ & 20.9249 & 6.5208 & 3.21 & 1.53 $\times 10^{-3}$ \\ 
  $Y \cdot M^3$ & -10.2445 & 2.2975 & -4.46 & 1.30 $\times 10^{-5}$ \\ 
  $Y \cdot Z_{\rm l}$ & 6.0335 & 3.6231 & 1.67 & 9.73 $\times 10^{-2}$ \\ 
  $Y \cdot Z_{\rm l}^2$ & 6.6911 & 1.8076 & 3.70 & 2.70 $\times 10^{-4}$ \\ 
  $Y \cdot Z_{\rm l}^3$ & 1.4383 & 0.3446 & 4.17 & 4.29 $\times 10^{-5}$ \\ 
  $Y \cdot M \cdot Z_{\rm l}$ & 10.5994 & 2.1842 & 4.85 & 2.28 $\times 10^{-6}$\\ 
  $Y \cdot M^2 \cdot Z_{\rm l}^2$ & 2.0149 & 0.3637 & 5.54 & 8.43 $\times 10^{-8}$ \\ 
   \hline
\end{tabular}
\end{table}

\begin{table}[ht]
\centering
\caption{As in Table \ref{table:mod-xm} but for $R_{\rm o}$ in
  Eq.~(\ref{eq:mod-ru}).} 
\label{table:mod-ru}
\begin{tabular}{rrrrr}
  \hline\hline
& Estimate & Std. error & $t$ value & $p$ value \\ 
  \hline
(Intercept) & 1.2736 & 0.1486 & 8.57 &  $< 2.0 \times 10^{-16}$ \\ 
  $M$ & -1.6732 & 0.4195 & -3.99 & 7.33 $\times 10^{-5}$ \\ 
  $M^2$ & 2.3650 & 0.4275 & 5.53 & 4.45 $\times 10^{-8}$ \\ 
  $M^3$ & -0.8356 & 0.1537 & -5.44 & 7.51 $\times 10^{-8}$ \\ 
  $Z_{\rm l}$ & 0.2009 & 0.0626 & 3.21 & 1.40 $\times 10^{-3}$ \\ 
  $Z_{\rm l}^2$ & 0.2366 & 0.0170 & 13.96 &  $< 2.0 \times 10^{-16}$ \\ 
  $Z_{\rm l}^3$ & 0.0479 & 0.0023 & 21.13 &  $< 2.0 \times 10^{-16}$ \\ 
  $Y$ & -1.1958 & 0.3826 & -3.13 & 1.85 $\times 10^{-3}$ \\ 
  $M \cdot Z_{\rm l}$ & -0.3292 & 0.0675 & -4.88 & 1.33 $\times 10^{-6}$ \\ 
  $Y \cdot M$ & 7.5938 & 1.2519 & 6.07 & 2.14 $\times 10^{-9}$ \\ 
  $Y \cdot M^2$ & -8.9430 & 1.3833 & -6.46 & 1.89 $\times 10^{-10}$ \\ 
  $Y \cdot Z_{\rm l}$  & 0.3827 & 0.0441 & 8.68 &  $< 2.0 \times 10^{-16}$ \\ 
  $M^2 \cdot Z_{\rm l}^2$ & -0.1145 & 0.0183 & -6.26 & 6.75 $\times 10^{-10}$ \\ 
  $Y \cdot M^3$ & 3.1766 & 0.5092 & 6.24 & 7.63 $\times 10^{-10}$ \\ 
  $M^3 \cdot Z_{\rm l}^3$ & -0.0109 & 0.0024 & -4.64 & 4.14 $\times 10^{-6}$ \\ 
  $Y \cdot M \cdot Z_{\rm l}$ & -0.2572 & 0.0483 & -5.33 & 1.33 $\times 10^{-7}$ \\ 
  $T_{169}$ & -1.7465 & 0.0184 & -95.08 &  $< 2.0 \times 10^{-16}$ \\ 
  $T_{169}^2$ & 0.3336 & 0.0047 & 71.08 &  $< 2.0 \times 10^{-16}$ \\ 
  $M \cdot T_{169}$ & 0.3018 & 0.0337 & 8.96 &  $< 2.0 \times 10^{-16}$ \\ 
  $Y \cdot T_{169}$ & 0.0704 & 0.0089 & 7.88 & 1.26 $\times 10^{-14}$ \\ 
  $M^2 \cdot T_{169}$ & -0.1263 & 0.0187 & -6.77 & 2.74 $\times 10^{-11}$ \\ 
   \hline
\end{tabular}
\end{table}

\subsection{Sensitivity to the uncertainties in stellar mass and metallicity estimates
}\label{sec:sensitivity} 

The statistical models described in the previous section allow 
determining the HZ and CHZ characteristics without the need to rely on
time-consuming stellar evolution computations, after the mass and the
metallicity of the host star are known. In the real world these quantities are
subject to measurement or estimation errors, which propagate into the model
output.  Thus, it is important to estimate how the current uncertainties
in the model input impact on the inferred CHZ boundaries.
To our knowledge, this analysis is still lacking in the literature; this is the
first attempt 
to quantitatively estimate the uncertainty propagation. 

More in detail we evaluated, by means of Monte Carlo simulations, the relative
variations on $R_{\rm i}$ and $R_{\rm o}$ induced by an uncertainty of 0.1~dex
on $\log Z$ and of 0.05~$M_{\sun}$ on the mass of the host star.  These values
are conservative estimates of the expected errors in [Fe/H] by spectroscopic
measurements and of those in the stellar mass estimation by means of grid techniques in
presence of asteroseimic constraints \citep[see e.g.][]{
  Quirion2010, Gai2011, Basu2012, Torres2012, griglia2013}. 

We define $(M_i, Z_i)$ as the vector composed by one of the 17 mass values
in the range [0.7 - 1.1]~$M_{\sun}$ with a step of 0.025~$M_{\sun}$ and
one of the 15 metallicity values in the range [0.005 - 0.04] with a step of
0.0025. We adopted in these computations a value of $\Delta Y/\Delta Z$ = 2.
Let $\sigma = (0.05, 0.1)$ be the vector of the uncertainties in mass
and log metallicity.  We sampled $10^5$ values of mass and metallicity
assuming a multivariate normal distribution with vector of mean $(M_i, \log
Z_i)$ and diagonal covariance matrix $\Sigma = {\rm diag} \; \sigma^2$.  For
each of these combination we evaluated $R_{\rm i}$ and $R_{\rm o}$ from the
statistical models described in Eq.~(\ref{eq:mod-rl}) and
(\ref{eq:mod-ru}). Due to the 
asymmetry of the obtained distributions, as a dispersion measure we computed
$\Delta R_{\rm i}$ and $\Delta R_{\rm o}$ as the differences between the 84-th
and the 16-th quantile of the two distributions. The relative variation is
obtained by the ratio of these values and that obtained for the
reference point $(M_i, Z_i)$.

\begin{figure*}
\centering 
\includegraphics[width=9cm,angle=-90]{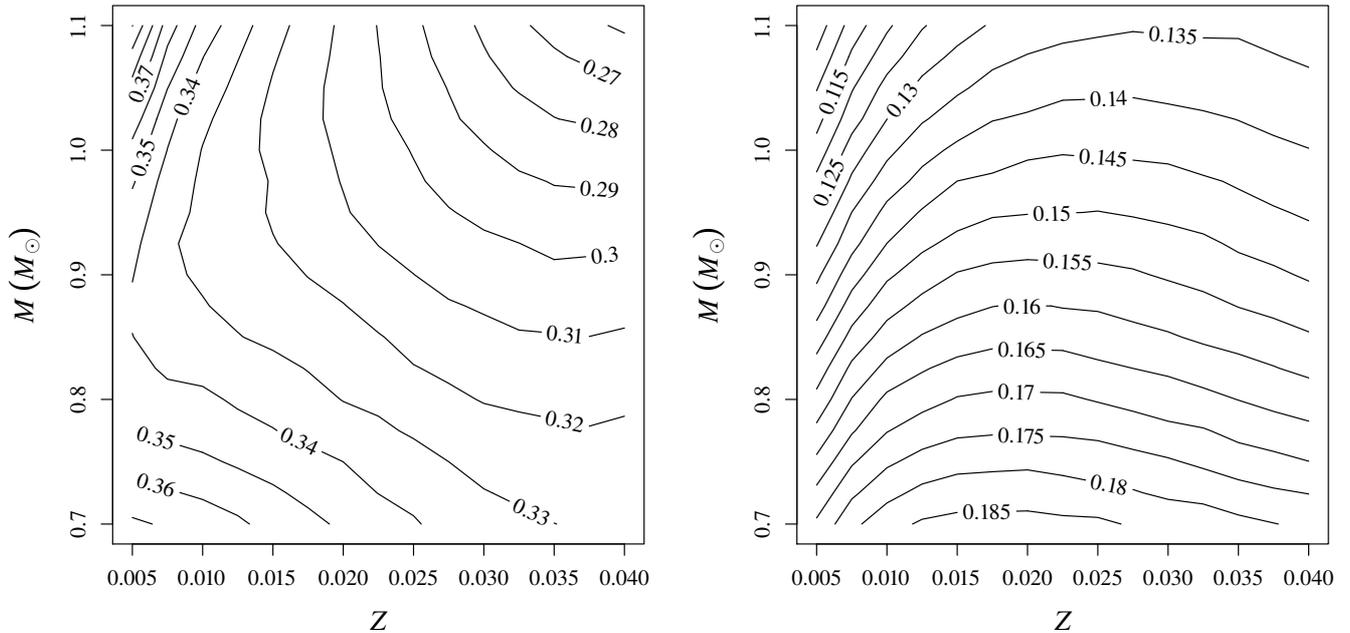}
\caption{(Left): contour plot of the relative variation $\Delta R$ in $R_{\rm
    i}$ due to  
  the simultaneous variation of the mass and the metallicity of the host star
  within their uncertainty ranges (see text for details). (Right): as in the
  left panel for the
  outer radius $R_{\rm o}$.  We adopted $\Delta Y/\Delta Z$ = 1.} 
\label{fig:perturded}
\end{figure*}

Figure~\ref{fig:perturded} displays the contour plot of the relative
variation on $R_{\rm i}$ (left) and  $R_{\rm o}$ (right) in dependence on the
mass and metallicity of the host star. The inner boundary of
the CHZ is most sensitive to the uncertainty of the model
input, with a relative variation ranging from about 27\% to 37\%, the higher
values occurring at
low metallicity. The relative variation of the outer boundary is
about one half of that of the inner boundary and it increases at low mass.

\begin{figure*}
\centering
\includegraphics[width=9cm,angle=-90]{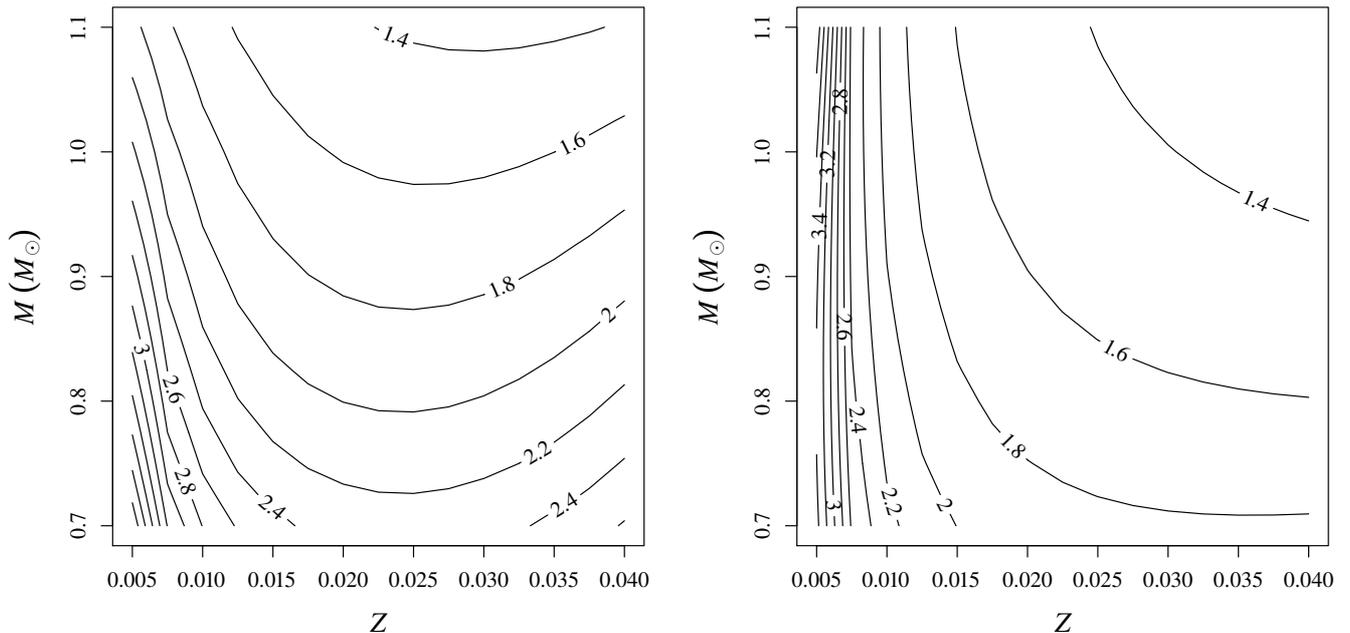}
\caption{(Left): contour plot of the ratio $(\Delta R_{\rm i, M}/\Delta
R_{\rm i, Z})$, i.e. the ratio of the variation of the inner boundary $R_{\rm
  i}$ due to the change of the host star mass to the one due to
  metallicity change.   
  Mass and metallicity are changed within their uncertainty
  ranges 
  (see text for details). (Right): as in the left panel, but for the
  outer radius $R_{\rm o}$. We adopted $\Delta Y/\Delta Z$ = 1.}
\label{fig:perturded-dmdz}
\end{figure*}

To separate the contributions to the relative variation due to the
uncertainties in mass and metallicity, we evaluated for each reference point
described above $\Delta R_{\rm i}$ and
$\Delta R_{\rm o}$ due to a variation of the mass $M_i \pm 0.05$~$M_{\sun}$
and of the 
metallicity  $\log Z_i \pm 0.1$~dex. Let $\Delta R_{\rm i, M}$ and $\Delta R_{\rm
  o, M}$ be the variation
in the inner and outer boundary due to the mass uncertainty, and  $\Delta
R_{\rm i, Z}$ and $\Delta R_{\rm 
  o, Z}$ the corresponding one due to the metallicity uncertainty.      
The left panel of Figure~\ref{fig:perturded-dmdz} shows the contour plot of
the ratio  $(\Delta R_{\rm i, M}/\Delta
R_{\rm i, Z})$ with respect to the mass and metallicity of the host
star. The right panel of the figure shows the corresponding quantities for the
outer boundary. The contribution of the mass uncertainty is always greater
than that due to the uncertainties in metallicity estimates and the difference increases towards low
masses and low metallicities. 

\section{Excluding the constant albedo assumption}
\label{sec:flux}

The results presented in the previous section mainly rely on the climatic
computations presented in \citet{Kasting1993}. 
These models have recently been revised by K13, who presented
an updated 
and improved climate model obtaining new estimates 
for HZ. For our purposes the most relevant difference is that in  
computating the
HZ boundaries we no longer assume a constant albedo at a given 
equilibrium temperature. The purpose of this section is to assess the
effect of this modification on the discussed HZ characteristics. 

To compute the incident flux the ``BT Settl'' grid of atmospheric models
\citep{Allard2007} has been adopted in K13. Solar metallicity was adopted 
in the computations (Kopparapu, 2014, private communication).

K13 reported a parametrisation of the $S_{\rm eff}$
in dependence on the effective temperature of the star,
for several explored HZ extensions. In the following we 
refer to their 
"moist greenhouse'' for the inner boundary
and "maximum greenhouse'' for the outer one.  

For inner and outer boundaries the critical effective flux is obtained by the
relations
\begin{equation} 
S_{\rm eff} = S_{\rm eff, \sun} + a T_* + b T_*^2 + c T_*^3 + d T_*^4 \; ,
\label{eq:seff}
\end{equation}
where $T_* = T_{\rm eff} - 5780$~K. The values of the coefficients $a,
b, c, d$ are given in Table~3 of  K13. 
The boundaries of the HZ are then obtained from Eq.~(\ref{eq:d-seff}).

\begin{figure*}
\centering
\includegraphics[width=8.5cm,angle=-90]{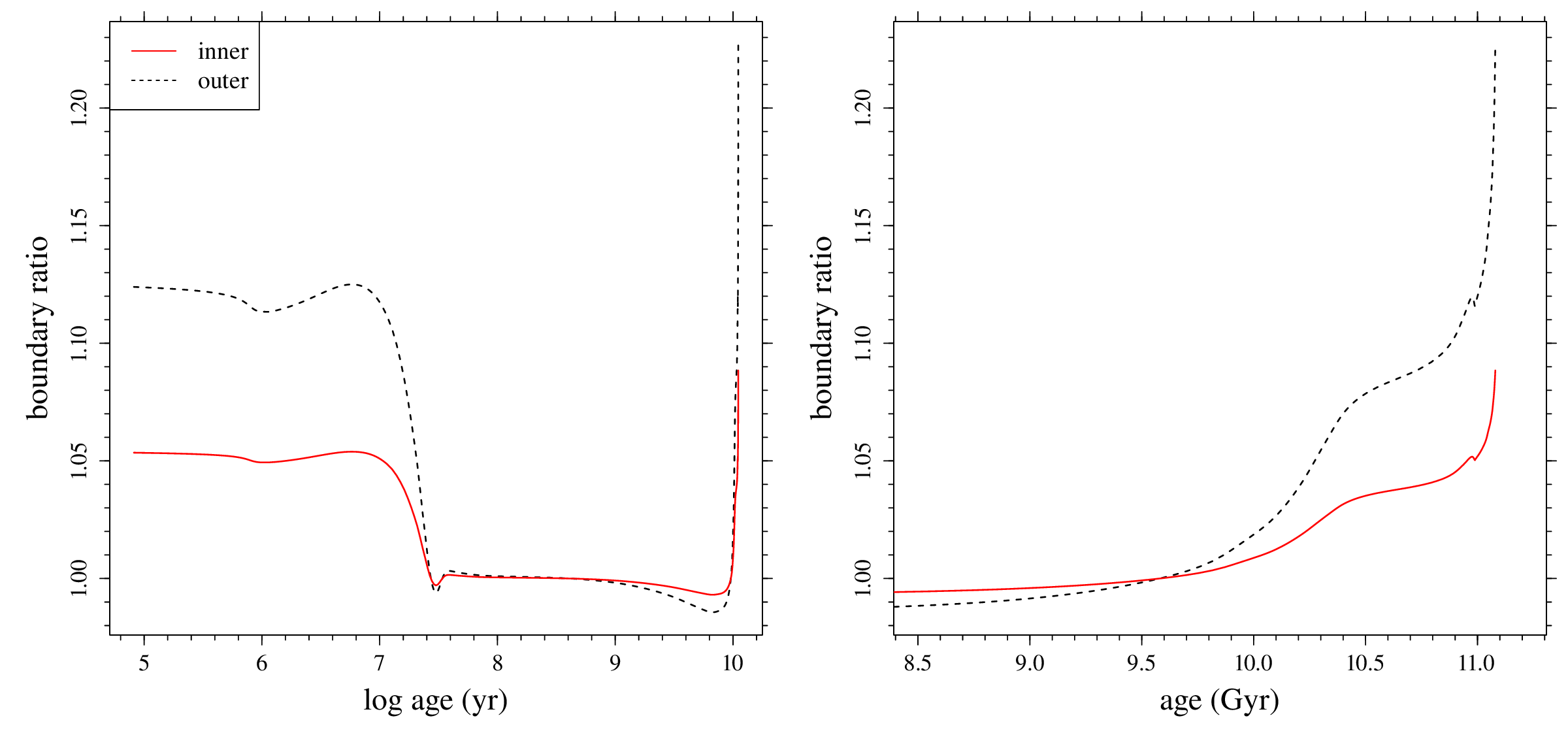}
\caption{(Left): ratio of the inner and outer boundaries for the solar model obtained
  from the new 
  climate models of \citet{Kopparapu2013} and the reference scenario. The 
  temperature adopted for the outer boundary computation is $T_{\rm o}$~=~203~K. 
(Right): detail of the red-giant branch evolutionary stage.
}
\label{fig:HZ-ratio-flux}
\end{figure*}

\begin{figure}
\centering
\includegraphics[width=9cm,angle=-90]{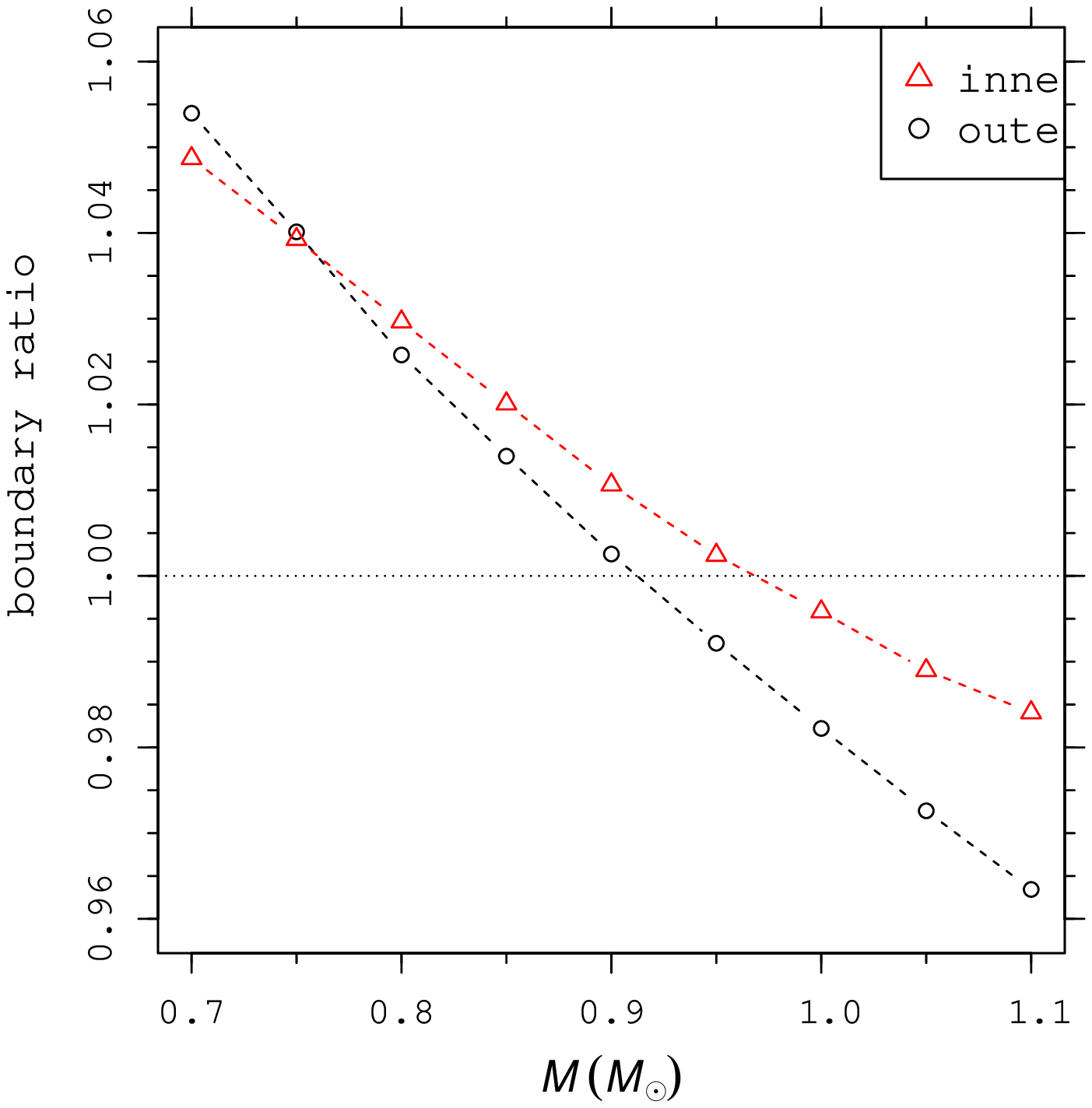}
\caption{Ratio of the inner and outer boundaries of the 4~Gyr CHZ obtained
  from the new 
  climate models of \citet{Kopparapu2013} and the reference scenario. The 
  temperature adopted for the outer boundary computation is $T_{\rm o}$~=~203~K. 
}
\label{fig:HZ-ratio-bound}
\end{figure}

With respect to our reference scenario, a systematic bias is expected on
  both 
 the inner and the outer boundaries of 
the HZ since the critical effective stellar fluxes for the solar model of
K13 are different from those adopted in Sect.~\ref{sec:results}. These biases can be
computed -- as resulting from Eq.~(\ref{eq:d-seff}) -- from the square root of
the ratio of the effective stellar fluxes of 
the two scenarios, and
are about 11\% and 7\% on the inner and outer boundary. 
The changes of these biases with time will reveal
the importance of taking into account the planetary albedo change during
stellar evolution. 

All these effects are shown in
Fig.~\ref{fig:HZ-ratio-flux} for a model of 1.0 $M_{\sun}$, $Z =
0.015$ (assuming $\Delta Y/\Delta Z$ = 1)\footnote{This set is the closest to 
our own standard solar model ($Z = 0.0137$, $Y = 0.253$) in the available grid
of stellar models.}. The figure shows the ratio of the inner and outer
boundaries obtained from Eq.~(\ref{eq:seff}) and
(\ref{eq:d-seff})
 with respect to those of
our reference scenario, corrected for the multiplicative systematic biases
due to the different solar critical fluxes adopted, as a function of the time.  
During the main sequence and sub-giant branch evolution the
ratios between the HZ boundary estimates are close to 1.0 (the largest difference
is smaller than 0.5\%).
As expected, the differences are larger during the pre-main sequence and
the red-giant branch (RGB) phases, where the approximation of constant albedo is
more problematic because of the large change in effective temperature with respect
to the solar one. The difference in the computations for the inner and the outer
boundary is
about 5\% and 12\% in the pre-main sequence of the stellar evolution,
and it can be as large as 
25\% for the outer boundary at the RGB tip.
A systematic shift in MS is present for stars of different mass, because of the
differences in their effective temperature. 
For a star of 0.7~$M_{\sun}$ the typical biases in MS are about 4.5\% and 10\% at
inner and outer  
boundary, while for a model of 1.1~$M_{\sun}$ they are $-1.2\%$ and $-2.5\%$.  

To evaluate the relevance of this effect, these values can be
  compared  with those  arising from  
the current uncertainty in the stellar observables affecting the estimates of
the  
quantities in Eq.~(\ref{eq:d-seff}). In fact, these uncertainties can lead to 
large variations on the actual HZ determination \citep[see e.g. the discussion
  in][]{Kaltenegger2011}.  
For the computation, we assumed 100 K as a conservative uncertainty in  $T_{\rm
  eff}$ 
\citep[see e.g.][]{Gai2011, Basu2012, griglia2013}. Regarding the intrinsic
luminosity, whenever no accurate distance are available (as for the majority
of the {\it Kepler} targets), an estimate of the uncertainty can be obtained
 from the Stefan-Boltzmann
relation $L = 4 \pi \sigma R^2 T_{\rm eff}^4$, where $R$ is the stellar radius and
$\sigma$ is the Stefan-Boltzmann 
constant. In presence of asteroseismic constraints a relative error on the
stellar radius of
about 2.5\% can be reached by means of grid-based techniques \citep{Basu2012,
  Mathur2012, griglia2013}. We adopted a 5\% relative error in radius as a 
  more conservative estimate, accounting for the variability arising from
  different 
estimation pipelines.
By simple error propagation it is then possible to estimate an uncertainty of
about 12\% on $L$ for solar effective temperature.
Then, the error propagation in Eq.~(\ref{eq:d-seff}) shows that the relative error
on the derived 
boundary distances, due only to the current 
observational uncertainty, is of about 6\%. 
In summary, excluding the constant albedo has -- on the HZ  characteristics
evaluated on this paper -- an influence that is lower than or equal to the
observational uncertainties, except for the outer boundary of our lowest mass model.

Moreover, the estimates of the HZ boundaries are affected by the  
uncertainties in the critical stellar fluxes resulting form different climate
models. 
The observational uncertainty evaluated above 
is quite relevant as long as the HZ inner edge distance is evaluated, 
while it could be negligible for the outer edge. As an example, in the former
case, for  
the solar model the estimate of the inner boundary of the HZ from 3D climate 
model of \citet{Leconte2013} is 0.95 AU, while that from the 1D climate
model  
presented in K13 is 0.99 AU, i.e. a relative variation of about 4\%. 
In contrast,  
for the HZ outer boundary, the differences among the critical fluxes
computed by climate models are still larger than the observational uncertainty
\citep[see e.g. the discussion 
  in][]{Selsis2007, Kaltenegger2011}.

The impact of excluding the constant albedo assumption on the 
HZ and CHZ characteristics discussed in Sect.~\ref{sec:results}
is expected to be of a few percent, since these characteristics 
  are 
mainly based on the main-sequence stages of stellar evolution.   
To verify this point we evaluated the impact of the K13 climate model on
the 4~Gyr CHZ boundaries for metallicity $Z = 
0.015$ (with $\Delta Y/\Delta Z$ = 1). 
The comparison of results for different $Z$ are less meaningful since
the correction of the effective flux does not take into account the
metallicity of the host star.  Although metallicity-related
modifications to the spectrum of the incoming radiation in the climate model
would introduce little modifications on the
parametrisation presented in Eq.~(\ref{eq:seff}) (Kopparapu, 2014, private
communication), we avoided this possible source of bias and restricted ourself
to solar metallicity.  
 
Figure~\ref{fig:HZ-ratio-bound}
shows the ratio of the inner and outer boundaries of the CHZ obtained 
by adopting the K13 parametrisation of the effective flux
and those presented in the previous section. 
The inner and outer estimates of the reference scenario are corrected for
  the 
multiplicative biases discussed above, to account for the
differences in the critical solar fluxes.  
For the model of 1.0 $M_{\sun}$ the differences in the estimated boundaries are
of about  
$-0.5$\% and $-2.0$\% for the inner and outer boundary.
The maximum differences are about 5.0\% and 5.5\% for inner and outer boundary
for the model of 0.70~$M_{\sun}$. At
1.1~$M_{\sun}$ the differences are about $-1.5$\% and $-3.5$\% for inner and
outer boundary. 
The range of variation due to excluding the constant albedo hypothesis are 
thus much lower than the half-range of variation due to
stellar mass and metallicity uncertainty presented in
Sect.~\ref{sec:sensitivity}.

\section{Conclusions and discussion}\label{sec:conclusions}

We studied the temporal evolution of the main quantities characterising the 
habitable zone of low-mass stars from pre-main sequence phase 
to the helium flash at the red-giant branch tip.

We computed a fine grid of detailed stellar models in the mass range [0.70 -
  1.10]~$M_{\sun}$,  
for different metallicities $Z$ (i.e. from 0.005 to 0.04), and different
initial helium abundances $Y$ ($\Delta Y/\Delta Z$~=~1, 2, and 3).

To define habitability characteristics we focused our analysis on some 
features, such as the distance $d_{\rm 
  m}$ (in AU) for which the duration of habitability is longest, the
corresponding duration $t_{\rm m}$ (in Gyr), the width $W$ (in AU) of the zone
for which the habitability lasts $t_{\rm m}/2$, and the integral $I$ (in AU
Gyr) of the transit function for transits longer than 4~Gyr.
We also evaluated the inner ($R_{\rm i}$) and outer ($R_{\rm o}$) boundaries of
the 4~Gyr continuously habitable zone, which are useful for planning a survey
for life signatures in exoplanet atmospheres.

The large set of computed stellar models allowed the fit of analytical linear
models for $d_{\rm m}$, $t_{\rm m}$, $W$, $R_{\rm i}$, and $R_{\rm o}$. 
We found that the relative
error of the analytical models on the estimated characteristics is of the
order of a few percent or lower.
Thus, these accurate analytical models allowed us to perform  
the first systematic 
study of the variability of the HZ boundaries position due to the uncertainty
in the estimates of the host star mass and metallicity, without the need to
compute a huge number of detailed, but time-consuming, stellar models. 

A C program to compute the HZ and CHZ characteristics and an interactive web
form are available
on-line\footnote{\url{http://astro.df.unipi.it/stellar-models/HZ/}}.   
The web interface also allows interpolating stellar tracks for the required
mass  
and chemical composition relying on our fine grid of stellar models. These
tracks  
can be useful for computing synthetic spectra of the incoming radiation to be 
coupled to detailed climate models.

As expected and already shown by \citet{Danchi2013}, the metallicity plays a 
relevant role in the HZ evolution: the
duration of habitability at a given distance is longer for metal-rich stars
than for  metal-poor ones. The
HZ and CHZ for high-metallicity stars are closer to the host star than those
 around low metallicity stars.
For the first time, we also studied the effect of initial helium abundance; the
increase of its value has an effect that is opposite to that of increasing metallicity.

The analytical models allowed us to evaluate the uncertainties on the 
derived boundaries of the CHZ due to the unavoidable errors on the mass and
metallicity estimates
of the host star. These values are in fact subject to observational or
estimation uncertainty that propagate in the model output. Assuming an
error of 0.05~$M_{\sun}$ and 0.1~dex in the mass and log
metallicity, we found that the relative variation of the inner boundary
position ranges from about 27\% to 37\%, while that for the outer boundary
is about one half of these values.

The initial helium content of the host star was found to affect the 
determination of the 4~Gyr CHZ boundaries in a non-negligible way, mainly the inner one. 
The helium variation considered in this paper, caused by varying $\Delta
Y/\Delta Z$ from  3 to 1, led to a relative variation of the position of
$R_{\rm i}$ between about 5\% for
$Z = 0.005$ to about 30\% for $Z = 0.04$. The relative variation of the outer
boundary $R_{\rm o}$ is less important and is generally lower than 10\%.

The results presented in this paper are the first study of the
variability of the CHZ boundary position due
to the uncertainty on the host star characteristics. This kind of variability
is substantial and is of the same order or greater than the systematic
uncertainty due to several other factors reported in recent
literature, such as the use of different stellar evolution computations that
adopt slightly different input physics, the
refinement of climatic models, or the adoption of an integrated  
approach in the definition of the HZ. 

The first of these effects was evaluated by a direct comparison  of the
results presented here with the only overlapping one obtained with a different
stellar  evolution 
code by \citet{Danchi2013}. The comparison showed a difference in the CHZ
boundary of less than 5\%, a value higher than the random 
component due to linear modelling of Sect.~\ref{sec:modelli}, but
much lower than the variability due to the uncertainties in the estimates of 
the mass and chemical composition of the host star. 

The impact of different climate models on the HZ calculations is more
difficult to assess since there are many parameters involved. However, some
conclusions can be stated.  The modifications to the climatic models of
\citet{Kasting1993} presented by K13 and excluding the constant albedo
assumption was evaluated by
computing, for solar metallicity, the boundaries of the 4~Gyr CHZ 
adopting the estimates provided in K13.  With respect to the results obtained
with our reference scenario we found a maximum variation of about 4\%
at 0.7 $M_{\sun}$ and $-3$\% at 1.1 $M_{\sun}$. These uncertainties are much
lower than those arising from the uncertainty in the estimates of stellar
mass and metallicity.  Moreover, 
K13 reported a comparison of their estimates with those
by \citet{Selsis2007}, who adopted another different approach
to the boundary correction. The differences are lower than 6\% for the inner
boundary and even lower for the outer one.

The quantification of the internal uncertainty due only to stellar evolution
was facilitated by the fact that the stellar models are based on a sound
theory, with a relatively small set of input and parameters and that the results
can be verified by comparison with large data sets.  This is not the case
for climatic or geophysics models of planets different from Earth, since these
models depend on planetary parameters that are impossible to estimate before a
planet has actually been discovered and characterised \citep[see e.g. the
  extensive discussion in][]{Selsis2007}.  
Several efforts have been made to shed light on the impact of some
potentially 
important effects due to planetary characteristics, such as the planetary
mass \citep{Kopparapu2014}, its rotation speed \citep{Yang2014}, its tidal
obliquity 
evolution, that is, the change of the angle between the 
planet rotational axis and the orbital plane normal \citep{Heller2011}.  
The impact of these planetary characteristics can be very strong: recent
computations by means of a three-dimensional general 
circulation model by
\citet{Yang2014} have shown that slowly rotating planets can maintain an
Earth-like climate at nearly twice the stellar flux as rapidly rotating 
ones. These planet-specific parameters 
can be correctly taken into account only after a planet characterisation.
More investigations that focus on
the quantification of the internal uncertainty of these approaches are needed
to assess the relative importance of the many source of uncertainty that
concur in the HZ definition.

The main interest in defining the habitable zone is the hope of identifying
an actually inhabited  world.
The key concept in this research area is to find a terrestrial exoplanet
atmosphere severely out 
of thermochemical redox equilibrium \citep[see e.g.][]{Lovelock1965,
  Seager2013, Seager2013ApJ},  
assuming that
life fundamental chemical reactions release
biosignature gases -- such as oxygen or ozone -- 
as metabolic process by-products. 
The results presented in this work, suggesting good host star targets for
the observational follow-up, 
 can be useful for the planning of 
life-signature detection in exoplanet atmospheres.

The different established exoplanet atmosphere observational methods  have
different requirements in 
terms of location of the planetary target. 
Direct imaging of an
exoplanet and its atmospheric characterisation impose severe
requirements in terms of spatial resolution and contrast \citep[see
  e.g.][for a review]{Seager2010}, and it is currently limited to
bright and massive planets orbiting far from the host star.

A second way to study an exoplanet atmosphere composition is exploitable for
transiting exoplanets, that is, exoplanets that cross the line of sight from Earth
to the host star \citep[see
  e.g.][and references therein]{Seager2010}. This allows characterising
the  
  flux spectrum of the planet. The likelihood of a transit configuration is higher for 
planets at a short distance from their host star. 
In the light of the results presented here, 
this is a favourable condition since the highest chances to find a planet in the
4~Gyr CHZ were found for low-mass, near-solar metallicity stars. In these
cases the CHZ is close to the host star.

\begin{acknowledgements}

We are grateful to our anonymous referees for many stimulating suggestions
and comments that 
were a great help to us in clarifying and improving the paper.
We warmly thanks Ravi Kumar Kopparapu for his exhaustive answer to
our request for information.
We thank Paolo Paolicchi for carefully reading the paper and for
useful comments.
This work has been supported by PRIN-MIUR 2010-2011 ({\em Chemical and dynamical evolution 
of the Milky Way and Local Group galaxies}, PI F. Matteucci), PRIN-INAF 2011 
 ({\em Tracing the formation and evolution of the Galactic Halo with VST}, PI
M. Marconi), and  PRIN-INAF 2011 
 ({\em The M4 Core Project with Hubble Space Telescope}, PI
L. Bedin). 

\end{acknowledgements}

\appendix
\section{Detailed analytical models}\label{sec:append-models}

For convenience, we report here the full form of the
analytical models described in Sect.~\ref{sec:modelli}.
As in the main text we define $Z_{\rm l} = \log Z$,  $M_{\rm l} = \log M$, $K_1
= \log \left(\sqrt{(1-A)/0.7} \; /T_{169}^2 \right)$, $T_{169} = T_{\rm
  o}/(169 \; {\rm K})$, $K_2 =
\log \sqrt{(1-A)/0.7}$.

\begin{eqnarray}
\log d_{\rm m} ({\rm AU}) &  = & K_1  -1.2410 \nonumber\\
&& + 1.5820 \; M -1.2290 \; M^2 + 0.4822 \; M^3 \nonumber \\ 
&& + 1.0790 \; Z_{\rm l} + 0.5470 \; Z_{\rm l}^2 + 0.0952 \; Z_{\rm l}^3 \nonumber \\
&&+ 1.0832 \; Y -1.0129 \; M \cdot Z_{\rm l} + 0.0051 \; Y \cdot Z_{\rm l}^3 \nonumber\\
&&-0.2799 \; M^2 \cdot \; Z_{\rm l}^2 + 0.8673 \; Y \cdot M^2\nonumber\\
&& -0.8351 \; Y \cdot M^3  -0.0285 \; M^3 \cdot Z_{\rm l}^3 \nonumber\\
& & -0.0232 \; Y \cdot M^3 \cdot Z_{\rm l}^3
\end{eqnarray}

\begin{eqnarray}
\log t_{\rm m} ({\rm Gyr}) & = & -34.2982 + 109.9736 \; M_{\rm l} + 40.7217 \frac{1}{M}\nonumber\\
&  & + 1.5098 \; Z_{\rm l} + 0.0665 \; Z_{\rm l}^2 -4.7661 \; Y \nonumber\\
&   & -3.1963 \; Y \cdot M_{\rm l} -0.8552 \; Y \cdot Z_{\rm l} + 77.5873 \; T_{169} \nonumber\\
&   &  -36.3864 \; T_{169}^2 -212.5445 \; M_{\rm l} \cdot T_{169} \nonumber\\
&   & + 98.7415 \; M_{\rm l} \cdot T_{169}^2 -76.4845 \frac{1}{M} \cdot T_{169} \nonumber\\
&   & + 35.5695 \frac{1}{M} \cdot T_{169}^2 -0.1967 \; Z_{\rm l} \cdot T_{169}^2 \nonumber\\
&  & + 0.2764 \; Z_{\rm l}^2 \cdot T_{169} -0.1871 \; Z_{\rm l}^2 \cdot T_{169}^2 \nonumber\\
&  & +0.3196 \; Y \cdot T_{169}^2 + 2.1374 \; Y \cdot M_{\rm l} \cdot T_{169}^2
\end{eqnarray}

\begin{eqnarray}
\log W ({\rm AU}) & = & K_2 + 0.6789\nonumber\\
& & -0.9764 \; M +  1.1740 \; M^2 \nonumber\\
& & + 0.3371 \; Z_{\rm l} + 0.4138 \; Z_{\rm l}^2 + 0.0942 \; Z_{\rm l}^3\nonumber\\
&  & -0.2347 \; Y + 2.1227 \; Y \cdot M -2.6487 \; T_{169}\nonumber\\
& & + 3.0859 \; M \cdot T_{169} -1.7081 \; M^2 \cdot T_{169} \nonumber\\
& & -0.0065  \; Z_{\rm l}^3 \cdot T_{169} + 1.6171 \; Y \cdot T_{169} \nonumber\\
& & -2.2411 \; Y \cdot M \cdot T_{169}
\end{eqnarray}

\begin{eqnarray}
\log R_{\rm i} ({\rm AU}) & = &  K_2 -1.1957\nonumber\\
&& -0.2712 \; M -5.2040 \; M^2 +2.7739 \; M^3\nonumber\\
&& + 0.7318 \; Z_{\rm l}  -0.9036 \; Z_{\rm l}^2  -0.2442  \; Z_{\rm l}^3\nonumber\\
&& +4.8948 \; Y -5.4295 \; M \cdot Z_{\rm l} -1.3059 \; M^2 \cdot Z_{\rm
  l}^2\nonumber\\
&& -0.1098 \; M^3 \cdot Z_{\rm l}^3 -3.6520 \; Y \cdot M\nonumber\\
&&  + 20.9249 Y \cdot
M^2 -10.2445 \; Y \cdot M^3\nonumber\\
&&  + 6.0335 \; Y \cdot Z_{\rm l} + 6.6911 \; Y \cdot
Z_{\rm l}^2\nonumber\\
&& + 1.4383 \; Y \cdot Z_{\rm l}^3 + 10.5994 \; Y \cdot M \cdot Z_{\rm
  l}\nonumber\\
&& +2.0149 \; Y \cdot M^2 \cdot Z_{\rm l}^2
\end{eqnarray}

\begin{eqnarray}
\log R_{\rm o} ({\rm AU}) & = & K_2 +  1.2736\nonumber\\
&& -1.6732 \; M +  2.3650 \; M^2  -0.8356 \; M^3\nonumber\\
&& +0.2009 \; Z_{\rm l} + 0.2366 \; Z_{\rm l}^2 + 0.0479 \; Z_{\rm l}^3
\nonumber\\
&& -1.1958 \; Y -0.3292 \; M \cdot Z_{\rm l} + 7.5938 \; Y \cdot M\nonumber\\
&& -8.9430 \; Y \cdot M^2 + 0.3827 \; Y \cdot Z_{\rm l}\nonumber\\
&& -0.1145 \; M^2 \cdot
Z_{\rm l}^2 +3.1766 \; Y \cdot M^3 \nonumber\\
&& -0.0109 \; M^3 \cdot Z_{\rm l}^3 -0.2572 \; Y \cdot M \cdot Z_{\rm l} \nonumber\\
&& -1.7465 \; T_{169} + 0.3336 \; T_{169}^2\nonumber\\
&& +0.3018 \; M \cdot
T_{169} +0.0704 \; Y \cdot T_{169} \nonumber\\
&& -0.1263 \; M^2 \cdot T_{169}
\end{eqnarray}

\bibliographystyle{aa}
\bibliography{biblio}

\end{document}